# Simulation of perovskite thin layer crystallization with varying evaporation rates

M. Majewski[a], S. Qiu[b], O. Ronsin[a], L. Lüer[b], V. M. Le Corre[b], T. Du*[b,c], C. J. Brabec[b,c], H.-J. Egelhaaf[b,c], J. Harting*[a,d]

Perovskite solar cells (PSC) are promising potential competitors to established photovoltaic technologies due to their superior efficiency and low-cost solution processability. However, the limited understanding of the crystallization behaviour hinders the technological transition from lab-scale cells to modules. In this work, advanced Phase Field (PF) simulations of solution-based film formation are used for the first time to obtain mechanistic and morphological information that is experimentally challenging to access. The well-known transition from a film with many pinholes, for a low evaporation rate, to a smooth film, for high evaporation rates, is recovered in simulation and experiment. The simulation results provide us with an unprecedented understanding of the crystallization process. They show that supersaturation and crystallization confinement effects determine the final morphology. The ratio of evaporation to crystallization rates turns out to be the key parameter driving the final morphology. Increasing this ratio is a robust design rule for obtaining high-quality films, which we expect to be valid independently of the material type.

## Introduction

Perovskite solar cells have become a very promising option for the 3rd generation of photovoltaics(1). Single-layer solar cells with perovskites as the absorber layer have now reached astonishing power conversion efficiencies (PCE) of over 26%(2). These efficiencies can be reached due to the excellent optoelectronic properties of this class of materials, which include a high absorption coefficient, high carrier diffusion lengths, tunability of the band gap, and high defect tolerance(3). An prominent issue that hinders the upscaling of perovskite photovoltaics is the efficiency gap between lab-scale solar cells and modules(4). One reason is the poor control of the morphology of the active layer which strongly impacts the device performance(5), owing to a deficiency in the comprehensive understanding of its morphology formation. Even for the most investigated system, methylammonium lead triiodide (MAPbI$_3$), the interplay of the physical mechanisms involved in the crystallization process for one-step solution processing, including evaporation, nucleation, growth, and mass transport, remains poorly understood(6). The goal is therefore to gain a better understanding of the processes taking place in a drying film, notably for deposition techniques that can be used for large-scale production and roll-to-roll processing to exploit the potential of fully printed perovskite solar cells at industrial scale(7).

It has been established that drying in ambient conditions of the precursor solution is usually not sufficient to get a high-quality film(8). A widely employed route to improve film quality is to increase the evaporation rate. This can be achieved by several methods like gas quenching(8), anti-solvents(9), or vacuum drying(10). The morphology information measured on the final film using scanning electron microscopy (SEM)(11), X-ray diffraction (XRD)(12), or photoluminescence (PL)(13) helps to judge the quality of the film. However, these techniques are insufficient to understand how the film has formed, which is needed and indispensable to rationally control the film formation and therefore its quality. The crystallization process can be investigated in-situ by grazing-incidence wide-angle X-ray scattering (GIWAXS)(14,15), UV-vis absorption(16,17), white light reflectance spectroscopy (WLR)(18), and PL (16,17). These techniques help to understand the chemical transition from the solution to the crystalline perovskite film. However, the information gained on the evolution of grain sizes, crystallinity, and crystal arrangement is rarely reported (19). For MAPbI$_3$ the picture looks as follows: Lead (II) iodide and Methylammonium Iodide are dissolved in polar solvents. For strongly coordinating solvents such as DMSO(20) the formation of solid-state intermediates (SSI) is observed during drying, which can result in a needle-like structure in the final state(21). For weakly binding solvents, like 2-ME(22), direct conversion, without SSI, is also reported(5,14).

To obtain a qualitative description of the crystallization of the perovskite films, one can refer to nucleation and growth models, including the Lamer model(23–25), the Johnson–Mehl–Avrami–Kolmogorov (JMAK) model(12,22), classical nucleation theory(24), Volmer-Weber growth and Frank-van der Merwe growth(6) or the evolution of Voronoi cells(26). However, all the aforementioned models are not able to predict the spatial organization of the crystalline film, namely the roughness of the dry film, the stacking of the crystals, and the amount of uncovered substrate, which all have a strong impact on device performance(27). As a result, to the best of our knowledge, no coherent description of the formation pathway of solution-processed perovskite films based on physical phenomena can be found in the existing literature. Therefore, a theoretical framework is needed that can explain the process-structure relationship in more detail. This can be achieved by Phase Field (PF) simulations, a powerful tool used to investigate the kinetics of thermodynamic phase transitions based on a continuum description with diffuse interfaces. The thermodynamics is described using a free energy functional, and the kinetic evolution is usually governed by the Allen-Cahn and Cahn-Hilliard equations. PF simulations can describe the phase change between liquid to solid(28,29) and/or spinodal decomposition(30) in multicomponent mixtures. The effect of evaporation(31) and hydrodynamics(32) can be included as well. PF simulations for printable photovoltaic systems were presented by Wodo(33,34) in the case of amorphous organic photovoltaic blends and by Michels(35) for crystallizing films. We recently developed a PF framework taking into account liquid-liquid demixing, crystallization, and hydrodynamic effects in drying films(36). This allowed us to successfully simulate the bulk-heterojunction formation in printed organic solar cells featuring crystalline materials(37). For OPV, important features of the film bulk-heterojunction morphology is the relative size of the donor and acceptor domains, might they be due to crystallization and/or liquid-liquid phase demixing. The picture for perovskite material is different, and the solar cell performance is related notably to roughness, pinholes, as well as crystal sizes and orientation. Now, a decisive, unique advantage of our theoretical framework regarding its application to perovskites is precisely the possibility to investigate the fundamental problems of pinholes and surface roughness of the dried film by tracking the displacement and deformation of the film-vapour interface(36).

In this paper, we use PF simulations coupled to in-situ and ex-situ measurements in order to understand the formation mechanisms of solution-processed perovskite films. We focus on $MAPbI_3$ layers cast from a 2ME-NMP solution using blade coating, a well-established model process for upscaling. Model parameters in the simulation are kept generic on purpose in order to obtain results that are valid not only for the specific material system and solution deposition technique investigated in the current work but also for other perovskite materials. The impact of the solvent evaporation rate (varied by gas quenching) on the final film morphology and its formation pathway is investigated experimentally. In parallel, our Phase Field model is used to simulate the morphology formation process. The roughness, the vertical stacking of the crystals, the occurrence of pinholes in the film, the time-dependent crystallinity, and the crystal sizes are extracted from the simulation data. The PF simulations are validated extensively against experiments. Then, the simulations are used to gain deep insights into the film formation process. For the first time, the precise film formation pathway is established, explaining the large variations in morphology for solution-processed perovskite depending on the process parameters. This proves that our framework can be used to understand the impact of processing conditions on the final morphology, making use of a sound physical basis. As a result, fully printed perovskite solar cells with improved performance were fabricated. Their JV curves are fitted with an open-source drift-diffusion model(38). From the drift-diffusion simulations, the main reasons for the observed changes in performance are extracted and correlated to the observed changes in morphology. We believe this methodology can be useful for minimizing the morphological gap and the efficiency gap from lab to fab.

The structure of the work is as follows: first, the simulation technique, the time-dependent evolution of the drying film during a typical simulation, and the experimental methods are described. Second, the results are presented in the main part of the paper: to start with, the model is validated using extensive comparison with experimental data. Then, the mechanisms driving the morphology formation are described and discussed in detail based on both in-situ measurements and simulation results. Afterwards, evidence is given that the ratio of evaporation to growth rate is the determining factor for the formation of different morphologies. Finally, the performance of the fully printed solar cells is presented, and the relationship with film morphology is discussed. Third, the work is summarized, and future research directions are proposed in the conclusion.

## Simulation procedure and experimental approach

The Phase Field model used here is a adaption of the Cahn-Hilliard Allen-Cahn Navier-Stokes multi-component model presented in (36). The system is modelled with three volume fractions (see Supporting Information 1): one field variable for the solute ($\varphi_1$, perovskite material), one for the solvents ($\varphi_2$), and one for the air ($\varphi_3$), which is a buffer material compensating solvent removal due to evaporation(36). Additionally, two order parameters define regions of crystalline perovskite ($\phi_c$) and vapor ($\phi_{air}$) phases. Finally, two additional fields $v$ and $P$ allow for tracking the velocity and pressure in the film, respectively. Using a single solute and a single crystalline phase to represent the perovskite formation is a strong simplification since the crystallization of perovskite involves sophisticated chemistry with the formation of several ion complexes and sometimes colloidal aggregates and/or solid-state precursor crystals(22). However, our focus is on the physics of nucleation and growth and their impact on the morphology formation. For this, we will show that we can gain very useful insights without considering the details of the solution chemistry. Note that in the presently investigated system, it has been shown that direct perovskite crystallization is dominant(22,39).

The Gibbs free energy $G_v$ accounts for the entropic mixing and enthalpic molecular interactions as described in the Flory-Huggins theory, surface energy for all the considered interfaces, and an energy of phase change from the liquid to the solid state featuring an energy barrier. The chosen form of the free energy ensures that crystallization can only take place through a nucleation and growth process (see Supporting Information 1.1 and 2.1).

The evolution of the volume fraction fields $\varphi_i$ is given by the advective Cahn Hilliard equation

$$\frac{\partial \varphi_i}{\partial t} + \boldsymbol{v}\nabla \varphi_i = \frac{v_0}{RT}\nabla\left[\sum_{j=1}^{2}\Lambda_{ij}\nabla(\mu_j - \mu_3)\right] \quad (1)$$

This is the generalized form of the advection-diffusion equation, where $\Lambda_{ij}$ are the symmetric Onsager mobility coefficients, which depend themselves on the composition and the phase state. $\mu_j - \mu_3$ is the exchange chemical potential evaluated from the functional derivatives of the free energy $G_v$, $R$ is the gas constant, $v_0$ is the molar volume of a lattice size as defined in the Flory-Huggins theory, and $T$ is the temperature.

The evolution of the crystalline order parameter $\phi_c$ is given by the stochastic advective Allen Cahn equation

$$\frac{\partial \phi_c}{\partial t} + \boldsymbol{v}\nabla \phi_c = -\frac{v_0}{RT}M_c\frac{\delta \Delta G_v}{\delta \phi_c} + \zeta_{AC} \quad (2)$$

where $M_c$ is the mobility coefficient of the liquid-crystal interface and $\zeta_{AC}$ is an uncorrelated Gaussian noise triggering nucleation.

The velocity field $\boldsymbol{v}$ is obtained by solving the fluid mechanics equations, whereby the convective flows are due to the capillary forces arising from the liquid-solid and liquid-vapor interfaces, which can be derived from the phase fields. Advection is responsible for the solid particle transport, but it has been checked that it has no impact on the nucleation and growth process itself, and therefore the final morphologies are quite insensitive to the solution viscosity (see Supporting Information 1.4). This allows to use a damped kinetics for convective transport, which is necessary to have tractable simulation times.

To handle the evaporation of the solvent the top of the simulation box is initialized with a layer of air above the drying film. During the simulation, an outflux $j^{z=z_{max}}$ of solvent is applied at the top of the simulation box ($z = z_{max}$):

$$j^{z=z_{max}} = \alpha\sqrt{\frac{v_0}{2\pi RT\rho}}P_0(\varphi_2^{vap} - \varphi_2^{\infty}) \quad (3)$$

This expression corresponds to the Hertz-Knudsen theory, where $\alpha$ is the evaporation-condensation coefficient, $P_0$ is a reference pressure, and $\varphi_2^{\infty} = P_2^{\infty}/P_0$, with $P_2^{\infty}$ being the solvent pressure in the environment. $\varphi_2^{vap}$ is the calculated volume fraction in the vapor resulting from the local liquid-vapor equilibrium at the film surface. The displacement of the film-vapor interface is described with an additional Allen-Cahn equation.

2D cross-sections of the film are simulated. Initially, the fluid film is assumed to be fully amorphous and perfectly mixed, and initialized with 20% volume fraction of solute. This corresponds roughly to 1.3 M MAPbI$_3$ and is well below the volume fraction needed for crystallization ('crystallization threshold', $\varphi_{crit}$, see Supporting Information 2.2).

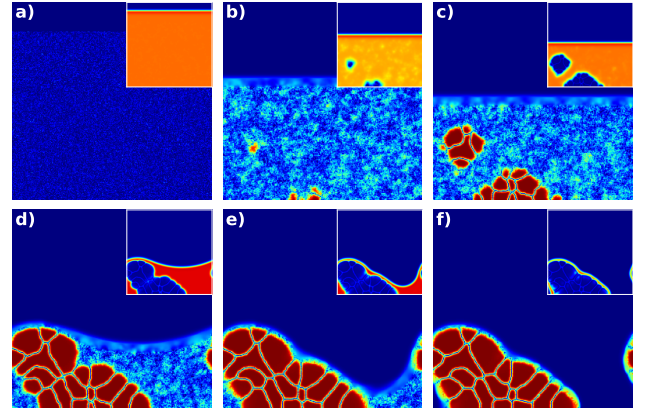

Figure 1: Time series of the simulated film drying for a low evaporation rate. The crystalline order parameter is shown as well as the volume fraction of the solvent (inset). The time increases from left to right and from top to bottom. (a) Initially, there is a homogenous, amorphous film without any crystals and a thin layer of air on top. (b) after some time, some solvent has evaporated, the height of the film is smaller and first nuclei appear. (c) In the supersaturation regime further nuclei appear. (d) When the volume fraction of solute decreases below the critical concentration $\varphi_{crit}$ no further nucleation can happen and only crystal growth takes place. (e) At a certain point, all the solute is consumed by the crystals, and only coarsening happens (note that the central crystal on the substrate is consumed by its neighbors from d) to e)). (f) Finally, the liquid film breaks, leaving pinholes. The full set of fields for this simulation is shown in Supporting Information 2.3.

Two sets of simulations are performed. Simulations of the first set solely differ in the evaporation rate of the solvent. In the second set only the crystallization rate is varied. While the evaporation rate is modified by adjusting the evaporation-condensation coefficient α (see equation 3), the crystallization rate can be adjusted by modifying the Allen-Cahn mobility $M_c$ (see equation 2). The effect of the annealing step in the experiment is mimicked by increasing the evaporation rate. A full list of parameters can be found in the Supporting Information 1.5. For each condition, five simulation runs are performed.

The behaviour of a typical simulation with a low evaporation rate is shown in Figure 1. The time increases from left to right and top to bottom. Initially, the condensed film is fully homogenous (Figure 1a). When the volume fraction of solute exceeds $\varphi_{crit}$ first nuclei form (Figure 1b). In the supersaturation regime nuclei keep appearing and all nuclei grow continuously (Figure 1c). In this example, the volume fraction of solute decreases in the liquid phase due to material consumption by crystallization. Below $\varphi_{crit}$ no further nucleation happens, and only growth proceeds (Figure 1d). When the amorphous phase reaches the thermodynamic equilibrium volume fraction of solute (saturation concentration), crystal growth terminates and the crystals coarsen (Figure 1e). Finally, the remaining solvent evaporates, and the substrate falls dry (Figure 1f).

Figure 2 shows a sketch of the experimental fabrication and characterization process. The precursors MAI and PbI$_2$, are dissolved in a solvent mixture of 2-methoxyethanol (2-ME) and N-Methylpyrrolidone (NMP). The solution is blade-coated onto an ITO substrate in ambient atmosphere. Gas quenching is applied at various air pressures immediately after deposition.

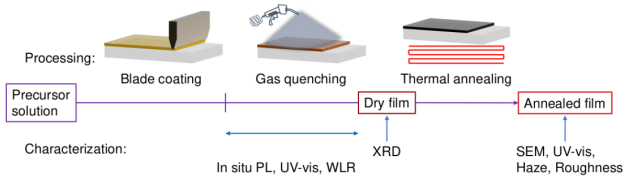

Figure 2: Schematic diagram of the experimental fabrication and characterization process. The precursor solution is blade-coated onto the substrate followed by gas quenching treatment at different air pressures. During gas quenching either a PL or a WLR signal is recorded. The XRD spectra are taken on the dry film before annealing. Further measurements are performed after annealing.

The advantage of gas quenching is that we only change the evaporation rate of the solvent, in contrast to temperature or solvent changes, which might also affect nucleation and growth kinetics and/or mechanisms. During gas quenching, in-situ PL and UV-vis, and WLR are recorded in parallel experiments using the same processing conditions. XRD diffractograms are taken on the dry film. Afterwards, the film is annealed on a hot plate. SEM and confocal images, as well as haze measurements, are performed after annealing to gain insight into the dry film morphology. The detailed descriptions of the setup, materials, and experimental procedures are reported in Supporting Information 2.4.

## Results

**Impact of the drying rate on the morphology and model validation**

The evaporation rate of a PbI2/MAI/2-ME/NMP film is controlled experimentally using gas quenching by varying the gas pressure from 0 to 2 bar. The evolution of the scaled film height $h/h_{fin}$ depending on the scaled time $t/\tau$, as measured by WLR, is shown in Figure 3 for three of the six tested gas pressures. Thereby, $h_{fin}$ is the expected average final film height and $\tau$ is the evaporation time for the fastest evaporation rate. In the simulations, the evaporation-condensation coefficient $\alpha$ is adjusted accordingly, such that there is a qualitative agreement of the evaporation rate variations between the simulations and the experiments (Figure 3a-b).

For all the evaporation rates the film height decreases at a constant evaporation rate through the drying process. For high evaporation rates, the solvent fully evaporates. At the end of the drying, the film has crystallized directly into MAPbI$_3$ as confirmed by the XRD spectra (Figure 3c). All films exhibit the perovskite phase originating from a diffraction peak located at 14.1°. For the 0 bar, the evaporation stops at a higher film height. This indicates that not all solvent is evaporated, but is trapped in the film (this hypothesis is supported by infrared reflectometry spectra, see Supporting Information 3.1). In the simulations, a similar behaviour is observed for low evaporation rates. There might be two reasons for solvent trapping in the film. First, it might be involved in SSI. Diffraction peaks located below 10° are present, which indicates that a second crystalline phase has formed. Note that this effect has not been taken into account in the simulation approach which focuses on direct crystallization only. Second, solvent might be trapped in small spaces in between or underneath the crystals, as evidenced in the simulations (see Supporting Information 3.2). In such a case, annealing enables further solvent removal (see Supporting Information 3.3).

A comparison of the simulated and experimentally observed morphologies after annealing is shown in Figure 4 and Figure 5. The amount of uncovered area of the sample and the final crystal sizes can be estimated from the SEM top view (Figure 4j-l). The vertical stacking of the crystals and the roughness of the sample can be obtained from the cross-section image (Figure 4m-o). For the simulation, these quantities can be calculated directly from the simulated fields in Figure 4g-i, (see Supporting Information 3.4). Note that for the quantification of pinholes and crystal sizes, data from simulation cross sections have been compared to data SEM top views (more statistics available): it is clear from the SEM observations that the deviations between crystal sizes and uncovered areas from cross sections and from top views are minimal, which allows such a comparison with minimal bias. The film morphologies after drying and annealing in the experiment and for five different simulation runs of each drying rate are shown in Supporting Information 3.5 and 3.6. The mechanisms of morphology formation will be discussed in detail, later in the manuscript.

The morphological variations show excellent agreement between experiments and simulations: the roughness (Figure 5a-b) of the film, the size of the crystals (Figure 5c-d), and the fraction of uncovered substrate (Figure 5e-f) decrease with increasing evaporation rate in both experiment and simulation. Furthermore, the order of magnitude of the scaled roughness of the film is comparable in the experiments and the simulations for all evaporation rates and Haze measurements (see Supporting Information 3.8) confirm the decreasing roughness of the samples. The trend of decreasing crystal sizes with increasing drying rate is recovered nicely in the simulations, although the values differ. The only exception to that is the slightly lower crystal size for naturally dried films, which is not recovered in the simulations. This might be due to the existence of the SSP phase for natural drying (see XRD spectra in Figure 3), which is not accounted for in the simulations.

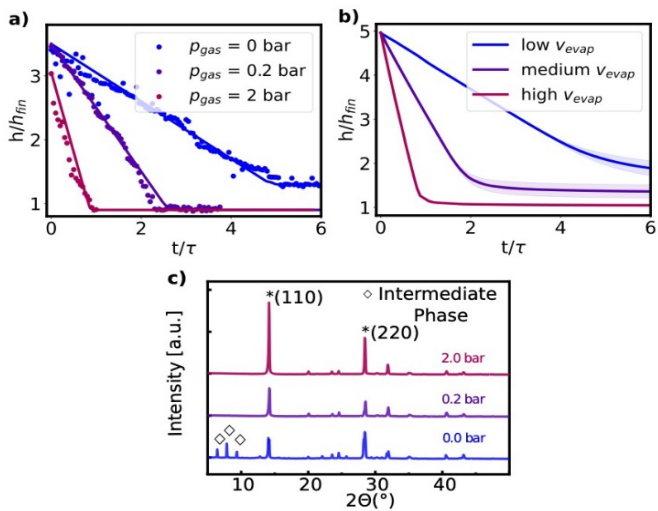

Figure 3: Drying film height in experiment (a) and in simulation (b). The lines in the experimental data are added as a guide for the eye. τ is the time until the film has reached its dry film height for the fastest evaporation rate. For the simulation, the shaded area indicates one standard deviation from the mean value. (c) XRD after drying.

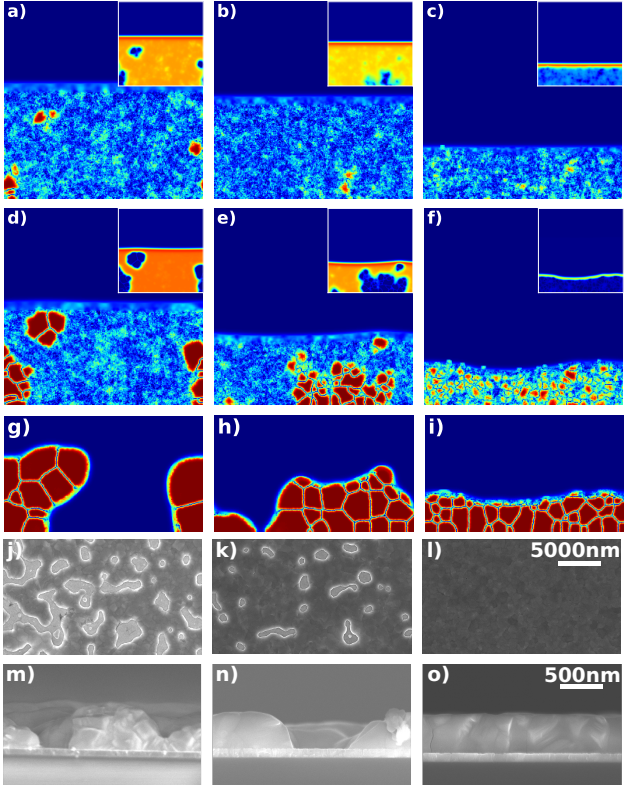

Figure 4: Crystalline order parameter and volume fraction of solvent (inset) for a low evaporation rate (a, d, g), a medium evaporation rate (b, e, h) and a high evaporation rate (c, f, i). The onset of crystallization (a, b, c), snapshot during drying with a similar crystallinity for all evaporation rates (d, e, f), and the morphology of the dry film (g, h, i, without annealing) are shown. SEM images for dry films before annealing obtained at different evaporation rates (j-l) top view, (m-o) cross-section. The snapshots correspond to the evaporation rates shown in Figure 3: (a, d, g, j, m) 0 bar and low $v_{evap}$, (b, e, h, k, n) 0.2 bar and medium $v_{evap}$, (c, f, i, l, o) 2 bar and high $v_{evap}$.

The fraction of uncovered substrate reaches its maximum value of roughly 25% for low evaporation rates for both experiment and simulation. Additionally, the relatively constant fraction of uncovered substrate and roughness for medium to high evaporation rates matches in simulation and experiment.

Overall, the observations in the experiments are very well captured by the simulation, including dramatic variations of the morphology features (roughness, coverage, crystal size increase) for very low evaporation rates and high-quality morphology (fully covered smooth films, stable crystal size) beyond a certain evaporation rate.

**Mechanisms driving the morphology formation**

In the following, the impact of the solvent evaporation rate on the crystallization pathway is analysed in detail with the help of the PF simulations and in-situ experimental data. The grain sizes calculated from in-situ PL data, the crystal sizes evaluated from the simulation, the UV/vis data together with the crystallinity curves from the simulations, and the simulated LaMer curves are shown in Figure 6. Since the simulated crystals rather correspond to the grains observed in SEM images, their size evolution differs from the crystallite sizes derived from the PL peak positions (see Supporting Information 3.9). Nevertheless, the time it takes for the crystals to appear can be compared between experiment and simulation. The trend of earlier crystallization onset for a higher evaporation rate is recovered in the simulations (Figure 6b). Regarding the crystallization rates, the trend of a higher crystallization rate for a higher evaporation rate is clearly visible in the simulations as well as from the UV-vis measurements (Figure 6d-e).

The reason for the different crystallization onset times is straightforward: Initially the volume fraction is below the volume fraction for the onset of crystallization $\varphi_{crit}$ (Figure 6e). The time required to reach that concentration is longer for a lower evaporation rate, so that crystallization starts later.

Once this critical concentration is reached, the decisive phase for the formation of the final morphology starts. From there, two distinct physical mechanisms drive the morphology formation in the simulations. A high evaporation rate leads to a high supersaturation and to a significant confinement of the enough to overcome the source term due to evaporation. As a result, the volume fraction of solute increases monotonously until the film has fully dried. In any case, for higher supersaturation, nucleation of the crystals is favored over crystal growth, which results in a larger number of smaller crystals in the final morphology.

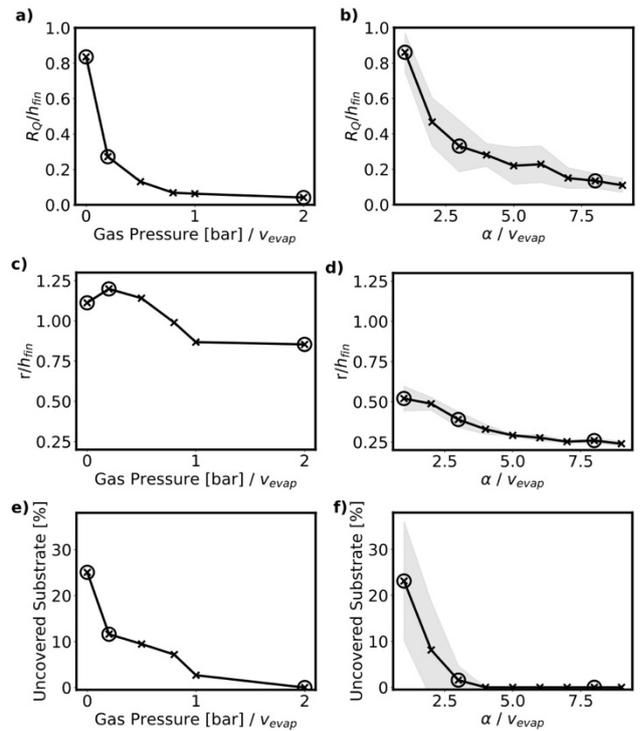

Figure 5: Different morphological properties of the perovskite films after annealing depending on the evaporation rate, obtained from experiment (left column) and simulation (right column). Evaporation rates for experiment and simulation are represented by the gas pressure of the air gun and the evaporation-condensation coefficient $\alpha$, respectively. From top to bottom: film roughness $R_Q$ normalized to the final film height $h_{fin}$ (a) and (b), mean crystal sizes normalized to the final film height (c) and (d), fraction of uncovered substrate (e) and (f). One standard deviation from the mean value is indicated as the shaded area. The encircled crosses are the experiments/simulations shown in Figure 3 and Figure 4 (A comparison between these morphological descriptors at the end of drying and after annealing in the simulation is shown in Supporting Information 3.7).

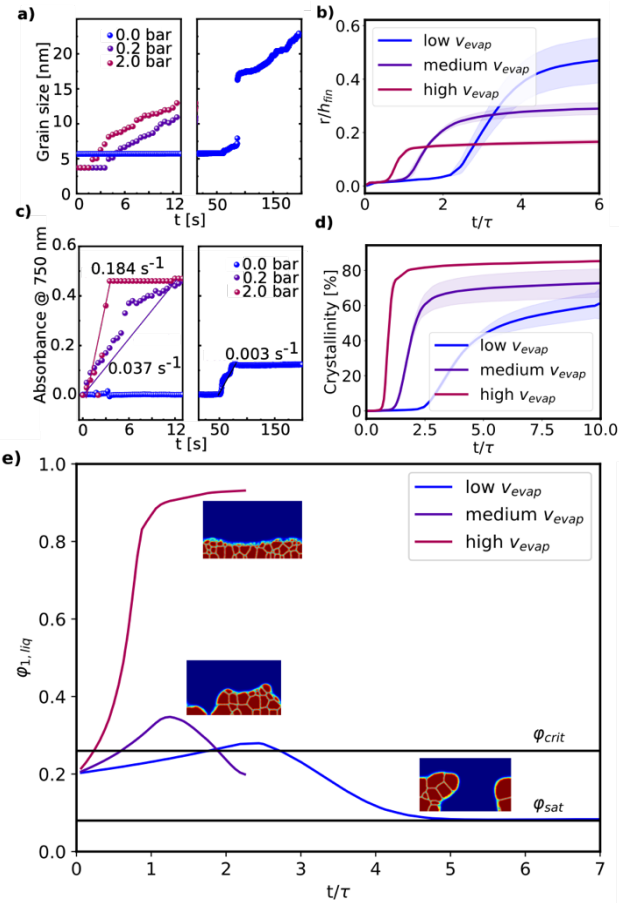

Figure 6: (a) Evolution of grain sizes during drying as calculated from PL spectra (see Supporting Information 3.9) (b) Evolution of relative grain sizes $r/h_{fin}$ as obtained from simulations. (c) Absorbance, measured at 750 nm, as a measure of the overall crystallinity of the film (see Supporting Information 3.9). (d) Crystallinity calculated from the simulations. (e) Simulated Lamer graphs: The volume fraction $\varphi_{1,liq}$ of solute in the condensed liquid film for the three selected evaporation rates is shown as a function of the evaporation time. The crystallization onset $\varphi_{crit}$, expected from the binary blend simulations, and the thermodynamic equilibrium volume fraction in the fluid phase $\varphi_{sat}$ (see Supporting Information 2.2) are also indicated (The evolution of the crystallinity, including annealing, is displayed in Supporting Information 3.3).

We now turn to the confinement effect: the space where the crystals can grow is more limited for high evaporation rates. This can be understood as follows: first, at the onset of crystallization (Figure 4a-c), the height of the liquid film is lower for higher evaporation rates, due to the delay between crossing the critical concentration $\varphi_{crit}$ and nucleation (see Figure 6e). During the whole process this trend persists for comparable amounts of crystallization (Figure 4d-f). Here again, for comparable amounts of crystalline materials, the film is thinner for higher evaporation rates. As a result, clusters with a size larger than the mean dry film height can only form for low evaporation rates (Figure 4d), while this is not possible when the evaporation rate is higher (Figure 4e-f). As a result, the dried film originating from a low evaporation rate shows a large roughness and pinholes (Figure 4g) whereas the films processed with a higher evaporation rate are smoother and pinhole-free (Figure 4h-i).

**The ratio of evaporation to crystallization rate determines the final film morphology**

We hypothesize that the effects of supersaturation and confinement on film morphology depend only on the balance space available for the crystals to grow and nucleate. This is drastically different for a low evaporation rate, where only a low supersaturation level is reached and where the crystals are allowed to evolve in a thick wet film with looser spatial constraints. The variations in the morphology when changing the evaporation rate can be explained by taking both effects into account.

We first focus on the supersaturation effect: After the saturation concentration is reached, nucleation is not instantaneous, so that a higher evaporation rate results in a higher supersaturation at the onset of crystallization. The level of attainable supersaturation for the different evaporation rates can be seen in Figure 6e, considering the maximum value reached by the solute volume fraction in the liquid $\varphi_{1,liq}$. The curves for low and medium evaporation rates are fully in line with the qualitative LaMer description in literature. Note that for a higher evaporation rate, crystallization is not fast enough to generate a solute sink term in the liquid film that is large between the rates of crystallization and evaporation. The natural and most simple descriptor for this balance is the ratio of both characteristic rates $v_{evap}/v_{cryst}$. To illustrate this, we vary the crystallization rate at fixed evaporation rate, by changing only the mobility $M_c$ in the Allen-Cahn equation (see equation 2). The corresponding dry film morphologies are shown in Figure 7. For a low crystallization rate (Figure 7a), a smooth film without pinholes is observed. The roughness increases with the crystallization rate and pinholes appear for the highest crystallization rate (Figure 7b-c).

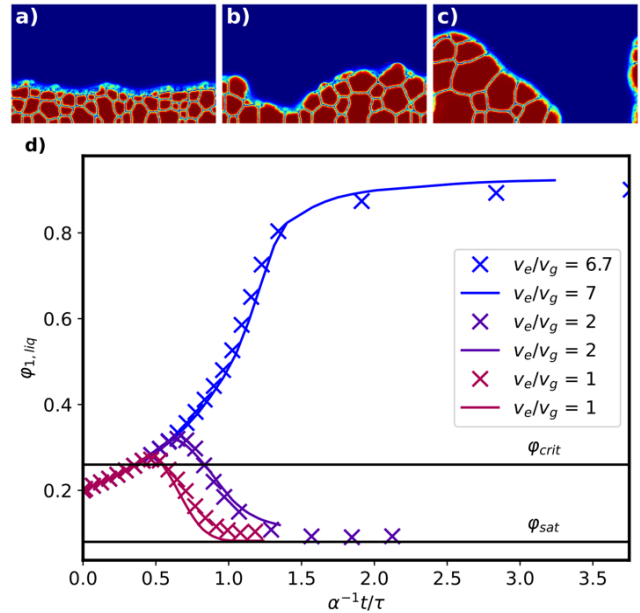

Figure 7: Film morphology after drying at fixed evaporation rate. (a) Slow crystallization rate. (b) Medium crystallization rate. (c) Fast crystallization rate. The film morphology in (c) is similar to a film with a slow evaporation rate regarding pinhole density, roughness, and crystal sizes. The morphology in (b) is like the one obtained for a medium evaporation rate and the one in (a) to a fast evaporation rate (see Figure 4). (d) LaMer curves of the simulations with variable crystallization rates at fixed evaporation rate (represented as crosses) and for simulations with variable evaporation rate at fixed growth rate (represented as lines, time scaled by the simulation input parameter α). The LaMer curves for similar ratios of evaporation to growth rate collapse to the same curves, which proves, that the ratio of these two quantities is the determining factor.

This behaviour is analogous to the morphology changes upon varying evaporation rate variation at fixed crystallization rate. Decreasing the crystallization rate has the same effect as increasing the evaporation rate. This is further illustrated in Figure 7d, where the LaMer curves obtained from crystallization rate variations (crosses) and evaporation rate variations (lines) are plotted together. This proves our hypothesis that dominantly the ratio of crystallization to evaporation rate controls the final film morphology.

The fact that fast evaporation is generally beneficial for perovskite film formation, independent of precursor composition and solvent system, further supports the hypothesis that it is mainly the ratio of both characteristic rates $v_{evap}/v_{cryst}$ which defines the film morphology, regardless of the atomistic, molecular, or chemical detail of the crystallization process. The good agreement between the experiments and the simulations presented in this work is also a strong indication that, except for very low evaporation rates (see Figure 3), intermediate species do not play a decisive role in perovskite film formation, at least for the solvent system used in our investigations.

From a more general perspective, this means that for any crystallizing material in a drying film, this effect is expected to be active as soon as the drying time and the crystallization time become comparable. A transition from smooth, pinhole-free films to rough films with incomplete substrate coverage should be experienced around $v_{evap} \sim v_{cryst}$. This can in principle be encountered for any fast-crystallizing material or even for slow-crystallizing materials if the process conditions are appropriate (very slow drying speed). Following this argument, we suggest that for any application experiencing rough films and/or pinholes, adjusting the processing condition by increasing $v_{evap}/v_{cryst}$ might lead to changes in the film morphology similar to the ones observed for perovskites.

**Dependence of the device performance on film morphology**

In this section, the relationship between the performance of the solar cells and the film morphology of the active layer is analyzed. For each evaporation rate, 20 fully printed solar cells with the layer stack glass/ITO/SnO2/MAPbI$_3$/P3HT/carbon are produced. Details on the stack can be found in the Supporting Information 2.4.

Guided by the PF simulations, the optimal performance can be achieved at an evaporation rate of 2 bar. Figure 8b shows the current density–voltage (J–V) curves of the best solar cell processed at 2 bar, which gives an efficiency of 19.34% with J$_{SC}$ of 23.10 mA/cm$^{-2}$, V$_{OC}$ of 1.06 V and FF of 79% obtained from the reverse scan but with a non-negligible hysteresis effect. To understand the impact of hysteresis, we measured the stabilized current and PCE. Thus, a stabilized PCE of 19.0% at maximum power point is achieved (see Supporting Information 4.1). The JV characteristics evaluated from the reverse scan are shown in Figure 8. While $V_{oc}$, FF, and PCE decrease significantly for lower evaporation rates, $J_{sc}$ decreases only slightly. For gas pressures of 0.5 bar and less, the JV-curves feature an S-shape. It should be noted that the yield of these devices follows the same trends: higher evaporation rates lead to higher yields (see Supporting Information 4.2).

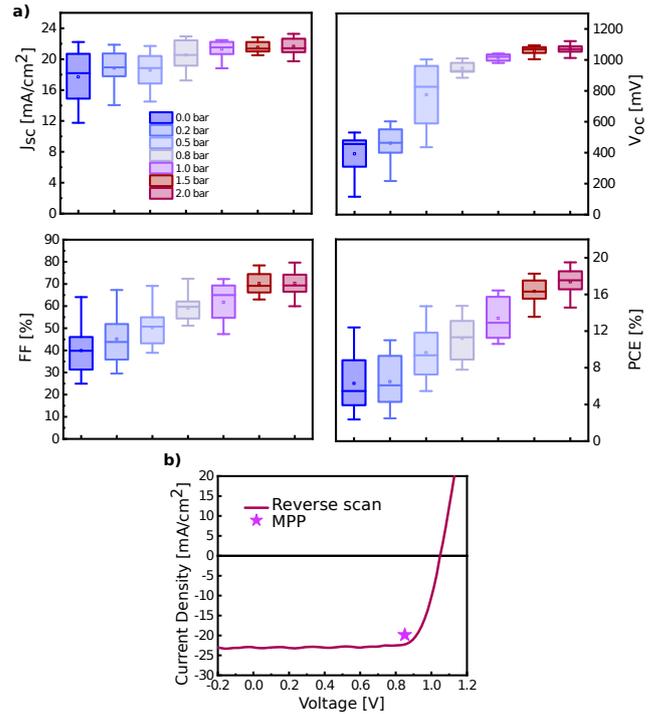

Figure 8: (a) Statistical photovoltaic parameters of short-circuit photocurrent density ($J_{sc}$), open-circuit voltage ($V_{oc}$), fill factor (FF), and power conversion efficiency (PCE). (b) J–V curves of a champion solar cell prepared with evaporation rate of 2 bar. (c) Time-dependent photocurrent and PCE at the MPP of the champion solar cell.

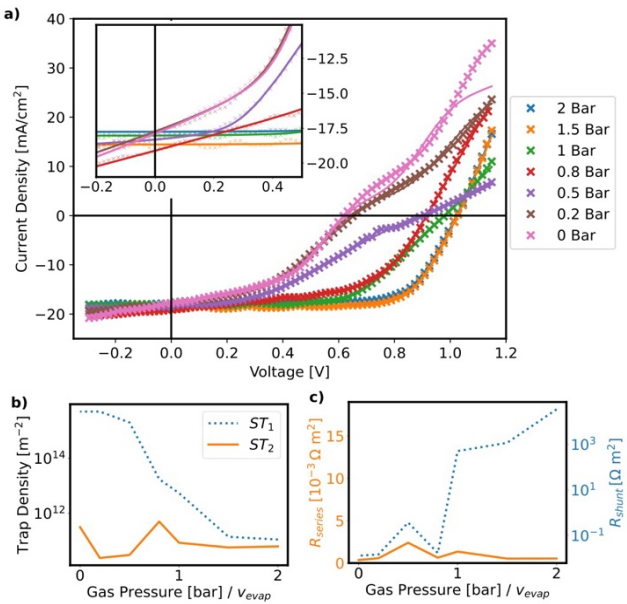

Figure 9: (a) JV curves fitted with a drift-diffusion model (experimental: crosses, fit: solid lines) and zoom around $J_{sc}$ (inset). (b), (c) Fitted parameters of the drift-diffusion model. (b) $ST_1$ and $ST_2$ are the interfacial trap densities between $SnO_2$ and the perovskite, P3HT and the perovskite respectively.

To gain a deeper understanding of this behaviour, the JV curves are fitted with an open-source drift-diffusion model (see Figure 9 and Supporting Information 4.3)(38). The fits indicate a decrease in shunt resistance and an increase in interfacial trap density along with a slight increase in series resistance upon decreasing the evaporation rate (see Figure 9c). The decrease of the shunt resistance is assigned to the increasing pinhole density as quantified by the fraction of the uncovered substrate area (see Figure 5), which leads to direct electrical contact between the hole and electron transport layers.

The increase of interface trap density obtained from the fits of the JV-curves is corroborated by the steady-state and time-resolved PL (TRPL) measurements, which indicate a substantial decrease of charge carrier lifetime with decreasing evaporation rate (see Supporting Information 4.4), caused by the increasing rate of nonradiative charge carrier recombination at the interfaces between perovskite film and charge transport layer. We tentatively ascribe the increase of interface trap density to the higher film roughness, which leads to larger interface areas, along with the deteriorated crystal quality as evidenced by the XRD spectra in Figure 3c and the SEM cross sections in Figure 4. Therefore, the deterioration of device performance with decreasing evaporation rate can be nicely understood from the trends in film morphology reported in the previous section.

## Conclusions

We investigated the effect of changing the solvent evaporation rate on the morphology formation of perovskite layers. MAPbI$_3$ cast from the solvent blend 2ME/NMP was used as a model system. The film height and the formation of the crystals were measured in-situ and the roughness, pinhole density, and crystal sizes of the final film were quantified. The evaporation rate was varied independently of any other effect using gas quenching. To gain insights into the morphology formation pathway, a recently developed simulation framework was used, and the simulations were successfully validated against experiments. A change from a rough film with many pinholes for low evaporation rates to a smooth, fully covered film, for high evaporation rates was observed. Furthermore, the change from larger crystal sizes at lower evaporation rates to smaller crystals at higher evaporation rates could be recovered with the simulations. A similar transition from smooth to pinhole-prone films was observed when changing the crystallization rate at a fixed evaporation rate. This indicates that the ratio of both rates is the main factor defining the final film morphology.

Our simulations allow to assign these changes to two effects: first, a high evaporation rate leads to high supersaturation and a high crystal density. Second, for a high supersaturation, these crystals nucleate at higher solute concentration, therefore in a thinner film, and thus in a much more confined space, leading to a smoother film. This is a purely geometrical effect. Both effects are expected to be valid independently of the specific chemical transitions involved. This transition from a rough film with many pinholes to a smooth, fully covered film is expected to occur in any crystallizing system where the evaporation rate and the crystallization rate are comparable independently of the atomic/molecular mechanisms responsible for the crystallization. The JV curves of the solar cells prepared from the different perovskite layers were fitted with a drift-diffusion model. The simulations suggest that shunts and interfacial trap density increase for decreasing evaporation rates.

Based on PF simulations of the fabrication process combined with device physics simulations, process-structure-property relationships can be established: in the present case, we provide the mechanistic explanation for the well-known fact that fast drying is necessary to get a pinhole-free film with low roughness, thus limiting trap state density and shunting, which is a prerequisite for achieving high PCE. Our simulations, thus, provide a so far unprecedented insight into the interplay between the drying and crystallization processes. They show that the key parameter determining the film morphology is the ratio of evaporation to crystallization rates. We expect this result to be valid for a wide range of materials, even beyond metal halide perovskites, as soon both rates lie in the same range.

In the future, we plan to understand and find the limits of the processing windows for high-quality perovskite films by varying the relevant process and material parameters. Adjusting, for example, the growth rate compared to the nucleation rate of the crystals is expected to allow the control of crystal sizes of the final film. In addition, the influence of the substrate on the film formation and the role of intermediate phases/solid state precursors can be investigated.

## Conflicts of interest

The authors declare no conflict of interest.

## Data availability

The simulation data and the scripts used to process it are publicly available at DOI: 10.5281/zenodo.10974044. The remaining data that support the findings of this study are included in the supporting information or available from the corresponding authors upon reasonable request.

## Acknowledgements

The authors thank Maxime Siber for helpful discussions. They acknowledge the financial support from the Deutsche Forschungsgemeinschaft (DFG) via the Perovskite SPP2196 programme (Project No. 506698391), the European Commission (H2020 Program, Project 101008701/EMERGE), the Helmholtz Association (SolarTAP Innovation Platform), ITRG 2495 (DFG), and the 'Solar Factory of the Future' as part of the Energy Campus Nürnberg (EnCN), which is supported by the Bavarian State Government (FKZ 20.2-3410.5-4-5).

V.M.L.C. acknowledge funding from the German Federal Ministry of Education and Research (BMBF) for the Solar TAP innovation platform under the Helmholtz Innovation Platforms funding line.

## Notes and references

# Simulation of perovskite thin layer crystallization with varying evaporation rates – supporting information

M. Majewski[1], S. Qiu[2], O. Ronsin[1], L. Lüer[2], V. M. Le Corre[2], T. Du[2,3], C. Brabec[2,3], H.-J. Egelhaaf[2,3], J. Harting[1,4]

[1]Helmholtz Institute Erlangen-Nürnberg for Renewable Energy (HIERN), Forschungszentrum Jülich GmbH

[2]Institute of Materials for Electronics and Energy Technology (i-MEET), Department of Materials Science and Engineering, Friedrich-Alexander-Universität Erlangen-Nürnberg, Erlangen, Germany

[3]Helmholtz Institute Erlangen-Nürnberg for Renewable Energy (HIERN), Immerwahr Straße 2, 91058 Erlangen, Germany

[4]Department of Chemical and Biological Engineering and Department of Physics, Friedrich-Alexander-Universität Erlangen-Nürnberg, Fürther Straße 248, 90429 Nürnberg, Germany

## Table of Contents

## 1. SI on the simulation model

This chapter presents the Phase Field model used in this work, which is a reduction of the multi-component model presented in[1]. In the current case, the system is modeled with three volume fractions: one field variable for the solute that can crystallize ($\varphi_1$, perovskite material), one for the solvents that can evaporate ($\varphi_2$), and one for the air ($\varphi_3$). Additionally, there are two order parameters. In the liquid phase, both order parameters are zero. In the crystalline phase, the crystalline order parameter $\phi_c$ is equal to one and the vapor order parameter is equal to zero. In the vapor phase, the vapor order parameter $\phi_{vap}$ is equal to one and the crystalline order parameter is equal to zero. Using a single solute and a single crystalline phase to represent the perovskite formation is a huge simplification since the crystallization of perovskite involves sophisticated chemistry with the formation of several ion complexes and sometimes colloidal aggregates and/or solid-state precursor crystals[2]. However, our focus is on the physics of nucleation and growth and their impact on the morphology formation. For this, we will show that we can gain very useful insights without taking into account the details of the solution chemistry. This is possible because, in the investigated system, it has been shown that direct perovskite crystallization is dominant, without SSI stage. In the simulation, the crystals nucleate spontaneously from random thermal fluctuations and may touch each other. To handle the interaction between impinging crystals, an additional labelling field $\theta$ is used. It is defined only in the crystalline phase, where the crystalline order parameter (and the volume fraction of solute) exceeds a certain threshold. Air is included in the system as a buffer material to be able to handle a deformable interface between the condensed phase and the vapor phase[1]. Finally, two additional fields $v$ and $P$ allow to track the velocity and pressure in the film, respectively.

## 1.1. Gibbs free energy

The energetic contributions of the system are collected in a free energy functional. This Gibbs Free energy $G$ can be split into a non-local and a local contribution

$$G = \int_V \Delta G_V dV = \int_V (\Delta G_V^{nonloc} + \Delta G_V^{loc}) dV. \tag{S1}$$

The non-local term $\Delta G_V^{nonloc}$ describes the surface tension arising from the various interfaces in the system and reads

$$\Delta G_V^{nonloc} = \sum_{i=1}^{3} \frac{\kappa_i}{2} (\nabla \varphi_i)^2 + \frac{\epsilon_{vap}^2}{2} (\nabla \phi_{vap})^2 + \frac{\epsilon_c^2}{2} (\nabla \phi_c)^2 + p(\phi_c) \frac{\pi \epsilon_g}{2} \delta_D(\nabla \theta), \tag{S2}$$

where $\kappa_i$ defines the strength of the surface tension related to the respective composition gradients, $\epsilon_{vap}$ defines the strength of the surface tension between vapor and non-vapor phase, $\epsilon_c$ defines the strength of the surface tension between crystalline and non-crystalline phases and $\epsilon_g$ defines the grain boundary energy. The term $\delta_D(\nabla \theta)$ equals one if there is a step in the marker field $\theta$ (at grain boundaries) and zero otherwise. Therefore, this term gives rise to an energy contribution at the interface between crystals, leading to the formation of boundaries between the crystals and enabling the handling of polycrystalline systems.

The local contribution $\Delta G_V^{loc}$ to the free energy can be written as

$$\begin{aligned}\Delta G_V^{loc} &= \left(1 - p(\phi_{vap})\right) \Delta G_v^{cond}(\varphi_i, \phi_c) + p(\phi_{vap}) \Delta G_v^{vap}(\phi_{vap}) + \Delta G_v^{crystvap}(\phi_c, \phi_{vap}) \\ &+ \Delta G_v^{num}(\varphi_i),\end{aligned} \tag{S3}$$

where the energy term for the condensed phase $\Delta G_v^{cond}(\phi_i, \phi_c)$ is written as

$$\begin{aligned}\Delta G_v^{cond}(\varphi_i, \phi_c) &= \rho_1 \varphi_1^2 \big(g(\phi_c) W + p(\phi_c) \Delta G_v^{cryst}\big) \\ &+ \frac{RT}{v_0} \left( \sum_{i=1}^{3} \varphi_i \ln(\varphi_i) + \sum_{i=1}^{3} \sum_{j>i}^{3} \varphi_i \varphi_j \chi_{ij,LL} + \sum_{j=2}^{3} \phi_c^2 \varphi_1 \varphi_j \chi_{1j,SL} \right).\end{aligned} \tag{S4}$$

The contribution inside the first brackets accounts for the change in energy density attributed to the change from the liquid to the solid phase. $g(\phi_c)$ and $p(\phi_c)$ are interpolation functions[3] chosen such that there is a higher energetical potential in the fluid phase than in the crystalline phase, and an energy barrier upon liquid-solid transition from $\phi_c = 0$ to $\phi_c = 1$. The following functions are used:

$$p(\phi_c) = \phi_c^2(3 - 2\phi_c) \tag{S5}$$

$$g(\phi_c) = \phi_c^2(1 - \phi_c)^2 \tag{S6}$$

$\rho$ is the density of the material, $W$ defines the height of the energy barrier between liquid and crystalline phase, $\Delta G_v^{cryst} = L_{fus}(T/T_m - 1)$ is the energy gain upon crystallization, whereby $T$ is the temperature, $T_m$ the melting temperature and $L_{fus}$ the enthalpy of fusion. The second part accounts for the entropic contribution and the enthalpic interactions between the different materials. $R$ is the gas constant, $v_0$ is the molar volume of a lattice size and $\chi_{ij,LL}$ is the Flory Huggins interaction parameter between amorphous materials $i$ and $j$. $\chi_{1j,SL}$ stands for the additional enthalpic interactions in the crystalline phase of the solute (material 1) between the crystalline solute and the solvent and/or air supposed to be amorphous. Note that in the binary blend considered here, only the solute material ('perovskite') can crystallize, and hence there are no $\chi_{ij,SS}$ parameters[4]. The values of $L_{fus}$, $W$, $\varepsilon_c$ are chosen so that the heat of fusion and surface tension are in a reasonable order of magnitude (75 *J/kg* and *60mJ/m²*, respectively), so that there is always an energy barrier to be overcome upon crystallization, and so that the interface is at least 6 grid nodes thick for proper space convergence.

The vapor phase is assumed to be an ideal mixture so that the energy contribution in the vapor phase can be written as

$$\Delta G_v^{vap}(\varphi_i) = \frac{RT}{v_0} \sum_{i=1}^{3} \varphi_i \ln\left(\frac{\varphi_i}{\varphi_{sat,i}}\right), \tag{S7}$$

where $\varphi_{sat,i} = P_{sat,i}/P_0$, with $P_{sat,i}$ being the vapor pressure and $P_0$ a reference pressure. The interaction between the crystalline and the vapor order parameter $\Delta G_v^{crystvap}(\varphi_1, \phi_c, \phi_{vap})$ reads

$$\Delta G_v^{crystvap}(\varphi_1, \phi_c, \phi_{vap}) = E(\varphi_1, \phi_c)\phi_c^2\phi_{vap}^2, \tag{S8}$$

with $E$ defining the strength of this interaction:

$$E(\varphi_1, \phi_k) = E_0 \frac{d_{sv}}{f(\varphi_1\phi_c, d_{sv}, c_{sv}, W_{sv})} \tag{S9}$$

where $E_0$ defines the interaction strength and $d_{sv}, c_{sv}, W_{sv}$ defining the strength, the center, and the width of the penalty function *f*:

$$\log(f(x, d, c, w)) = \frac{1}{2}\log(d)\left(1 + \tanh(w(x - c))\right) \tag{S10}$$

with *d, c,* and *w* defining the strength, the center and the width of the penalty. This contribution is added to prevent the vapor phase from penetrating into the crystalline phase and vice versa. This helps to ensure the stability of the crystals at the solid-vapor interface[3]. Finally, the purely numerical contribution $\Delta G_v^{num}(\varphi_i)$ ensures that the volume fractions stay in the range ]0,1[.

$$\Delta G_v^{num}(\varphi_i) = \sum_{i=1}^{3} \frac{\beta}{\varphi_i} \tag{S11}$$

The coefficient $\beta$ is chosen as small as possible to have the least possible impact on the thermodynamic properties and nevertheless provide numerical stability.

## 1.2. Cahn Hilliard & stochastic Allen Cahn equation

The evolution of the volume fraction fields is given by the advective Cahn Hilliard equation

$$\frac{\partial \varphi_i}{\partial t} + \boldsymbol{v}\nabla \varphi_i = \frac{v_0}{RT} \nabla \left[ \sum_{j=1}^{2} \Lambda_{ij} \nabla(\mu_j - \mu_3) \right] \tag{S12}$$

This is the generalized form of the advection-diffusion equation, where $\Lambda_{ij}$ are the symmetric Onsager mobility coefficients, which depend themselves on the composition and the phase state. They are typically lower in the crystalline phase and higher if a large volume fraction of solvent is present[3]. The Cahn Hilliard mobility coefficients are expressed as:

$$\Lambda_{ii}^{cond} = \omega_i \left(1 - \frac{\omega_i}{\sum_{k=1}^{n} \omega_k}\right) \tag{S13}$$

$$\Lambda_{ij}^{cond} = -\frac{\omega_i \omega_j}{\sum_{k=1}^{n} \omega_k} \tag{S14}$$

with

$$\omega_i = N_i \varphi_i f(\phi_c, d_{sl}, w_{sl}, c_{sl}) \prod_{j=1}^{n} \left(D_{s,i}^{\varphi_j \to 1}\right)^{\varphi_j} \tag{S15}$$

where $D_{s,i}^{\varphi_j \to 1}$ is the self-diffusion coefficient in the pure material. $\mu_j - \mu_3$ is the exchange chemical potential evaluated from the functional derivatives of the free energy $G$:

$$\mu_j - \mu_3 = \frac{\delta G}{\delta \varphi_j} - \frac{\delta G}{\delta \varphi_3} = \frac{\partial \Delta G_V}{\partial \varphi_j} + \nabla\left(\frac{\partial \Delta G_V}{\partial \nabla \varphi_j}\right) - \frac{\partial \Delta G_V}{\partial \varphi_3} - \nabla\left(\frac{\partial \Delta G_V}{\partial \nabla \varphi_3}\right) \tag{S16}$$

Nucleation, growth, coarsening and impingement of the crystals are described by the dynamic evolution of the crystalline order parameter based on the stochastic advective Allen Cahn equation:

$$\frac{\partial \phi_c}{\partial t} + \boldsymbol{v}\nabla \phi_c = -\frac{v_0}{RT} M_c \frac{\delta \Delta G_v}{\delta \phi_c} + \zeta_{AC} \tag{S17}$$

where $M_c$ is the mobility coefficient of the solid-liquid interface and $\zeta_{AC}$ is an uncorrelated gaussian noise with zero mean and a standard deviation of

$$\langle \zeta_{AC}(x,t), \zeta_{AC}(x',t') \rangle = \frac{2 v_0}{N_a} M_c \delta(t-t') \delta(x-x') \tag{S18}$$

where $N_a$ is the Avogadro Number. In the simulations presented in this work, since liquid-liquid demixing is not possible, the fluctuations on the crystalline order parameter are dominant for triggering nucleation. This is the reason why fluctuations have not been taken into account in the Cahn-Hilliard equation. Nevertheless, it was checked that including fluctuations on the volume fraction fields does not change the crystallization behavior.

## 1.3. Evaporation

The top of the simulation box is initialized with a layer of air above the drying film. To simulate the evaporation of the solvent, an outflux $j^{z=z_{max}}$ of solvent is applied at the top of the simulation box ($z = z_{max}$):

$$j^{z=z_{max}} = \alpha \sqrt{\frac{v_0}{2\pi RT\rho}} P_0 \left( \varphi_2^{vap} - \varphi_2^\infty \right) \tag{S19}$$

This expression corresponds to the Hertz-Knudsen theory[5–7], where $\alpha$ is the evaporation-condensation coefficient, $P_0$ is a reference pressure, and $\varphi_i^\infty = P_i^\infty / P_0$, with $P_i^\infty$ being the solvent pressure in the environment. $\varphi_i^{vap}$ is the calculated volume fraction in the vapor resulting from the local liquid-vapor equilibrium at the film surface.

The evolution of the vapor order parameter $\phi_{vap}$ is governed by the advective Allen Cahn equation for the vapor phase

$$\frac{\partial \phi_{vap}}{\partial t} + \boldsymbol{v} \nabla \phi_{vap} = -\frac{v_0}{RT} M_{vap} \frac{\delta \Delta G_v}{\delta \phi_{vap}} \tag{S20}$$

where $M_{vap}$ is the Allen Cahn mobility of the liquid-vapor interface. $M_{vap}$ is chosen high enough to ensure that the liquid-vapor equilibrium is maintained locally during the whole simulation time. Note that the air has also to be attributed a vapor pressure. Its value has no physical meaning, but has to be set to an (unrealistically) high value in order to prevent the air from penetrating the condensed phase.[8] The diffusive constant of the solute in the vapor phase is chosen small enough such that diffusive mass transport of separated domains through the vapor phase is prohibited. The diffusive constants of the solvent and the air in the vapor phase are chosen large enough to ensure a constant volume fraction of solvent through the whole vapor phase. Under these conditions, it is possible to obtain the correct drying kinetics by setting the outflux only at the top of the simulation box and not directly at the liquid-vapor interface[8]. This allows to have deformable liquid-vapor interfaces in the system and to obtain a rough film, even with pinholes.

## 1.4. Fluid dynamics

The equations of fluid dynamics are used to calculate the velocity field $\boldsymbol{v}$ which results from capillary forces that arise at all interfaces and thus resolve advective mass transport. Advective mass transport is responsible for crystals at the film surface being pushed downwards during evaporation due to liquid-vapor surface tension, and for neighbouring crystals to be attracted as a result of the flow field between them. At the here relevant system scales, the Reynolds number is always small, and fluid inertia can be neglected. Also, gravity can be neglected when compared to the resulting capillary forces generated. The fluid flow is assumed to be incompressible and can be described by a single velocity field[3]. As a result, the continuity equation reads:

$$\nabla \, v = 0 \tag{S21}$$

and the momentum conservation equation can be written as

$$-\nabla P + \nabla(2\eta_{mix} S) + F_\varphi + F_\phi = 0, \tag{S22}$$

where *P* is the pressure field ensuring incompressibility, which is a result of the resolution of the fluid mechanics equations, but has no relevant physical meaning for the current investigations and will not be discussed in any detail in the following. $S$ is the strain rate tensor $\eta_{mix}$ is the composition- and phase-dependent viscosity[3]:

$$\frac{1}{\eta_{mix}} = f\big(\delta_D(\theta_k)\phi_c\varphi_1, d_\eta, c_\eta, w_\eta\big) \sum_{i=1}^{3} \frac{\phi_i}{\eta_i} \tag{S23}$$

where $\eta_i$ is the viscosity of material *i*, $\delta_D(\theta_k)$ is one if the orientation parameter is present and zero otherwise, $d_\eta, c_\eta, w_\eta$ defining the strength, the center, and the width of the penalty function. The capillary forces arising from the volume fraction and order parameter fields can be written as[9]:

$$F_\varphi = \nabla\left[\sum_{i=1}^{3} \kappa_i(|\nabla\varphi_i|^2 I - \nabla\varphi_i \times \nabla\varphi_i)\right] \tag{S24}$$

and

$$F_\phi = \nabla\left[\epsilon_c^2(|\nabla\phi_c|^2 I - \nabla\phi_c \times \nabla\phi_c) + \epsilon_{vap}^2\left(|\nabla\phi_{vap}|^2 I - \nabla\phi_{vap} \times \nabla\phi_{vap}\right)\right] \tag{S25}$$

where *I* is the unit tensor.

In the present model, the chosen viscosity values are unrealistically high. Unfortunately, this is necessary to allow tractable simulation runtimes. However, we expect this assumption to have no impact on the transport processes at stake in the drying film, basing on following considerations. Crystallization starts in a solution at volume fractions of typically 20-30%, when the solution viscosity is somewhat larger and the diffusion coefficients smaller than in pure solvents, but let's assume $5 \cdot 10^{-3} Pa \cdot s$ and $10^{-9} m^2 \cdot s^{-1}$ as reasonable values. The time scale of the processes at stake (drying, crystallization) are in the range of $1 - 10s$. On the one hand, the characteristic diffusion length is as large as $L \approx \sqrt{Dt} \approx 10\mu m$, which ensures that the composition of the liquid solution remains fully homogeneous. On the other hand, crystal transport in this solution might happen through convection and/or diffusion. Using the Stokes-Einstein relationship, the diffusion coefficient of a $10nm$ crystal in the solution is expected to be about $D = 4 \cdot 10^{-12} m^2 \cdot s^{-1}$, to be compared with the kinematic viscosity of about $10^{-6} m^2 \cdot s^{-1}$. This suggest that convection is the dominant transport mechanism for the nanocrystals. To summarize, the situation to be simulated is that crystal transport occurs by convection in an homogeneous liquid solution, crystal diffusion being a negligible mechanism at first order. This is still the case in our model, even though the viscosities are unrealistically high and decoupled from the diffusion coefficients values: the use of realistic diffusion coefficients ensures homogeneity of the solution, and the fluid mechanics ensures convective crystal transport. Thereby, even with the very high viscosity values chosen, capillary forces from interfaces with typical surface tensions $\sigma = 20 mN \cdot m^{-1}$ can generate flows with velocities around $100 nm \cdot s^{-1}$. This is sufficient for crystal transport considering that the system's characteristic length is a few hundredth of nanometers and and the drying characteritic time around $10s$. Of course, crystal *transport* is underestimated, which might impact the final crystal spatial arrangement in the dry film, and therefore the exact value of roughness and pinhole density. However, crystal *formation* is not affected: it has been checked that crystal nucleation and growth processes are insensitive to the convective flows (see Figure S1), as expected in a homogeneous solution. Therefore the processes (confinement, supersaturation) responsible for the changes of morphology with drying rate (increase of roughness and pinholes density, decrease of crystal size with decrease of drying rate) are not impacted by the magnitude of the convective flows, which explains the qualitative robustness of our findings (see Figure S2).

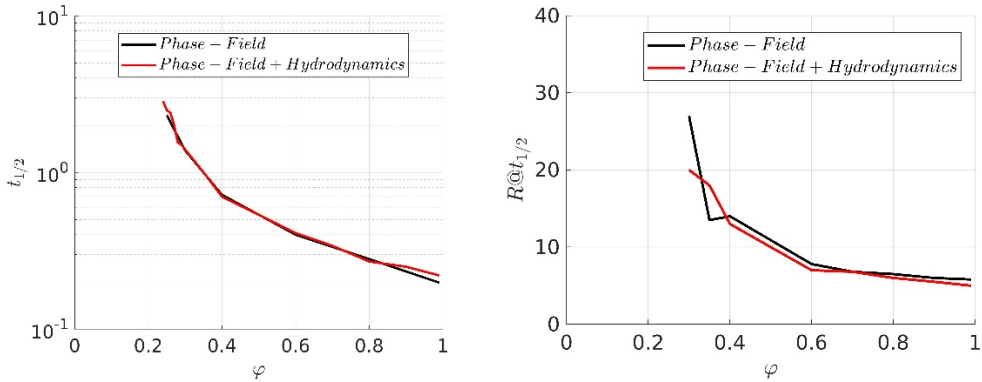

*Figure S 1: Crystallization half time $t_{1/2}$ (left) and crystal radius at $t_{1/2}$ (right) measured in simulations of binary solute-solvent blends at fixed solute volume fraction $\varphi$, without convective flows (black curve) and with convective flows (red curve). There is no impact of convective flows in the crystal nucleation and growth process. Note that below $\varphi = 0.4$, the number of crystals is very low and therefore the precision on the radius evaluation very poor.*

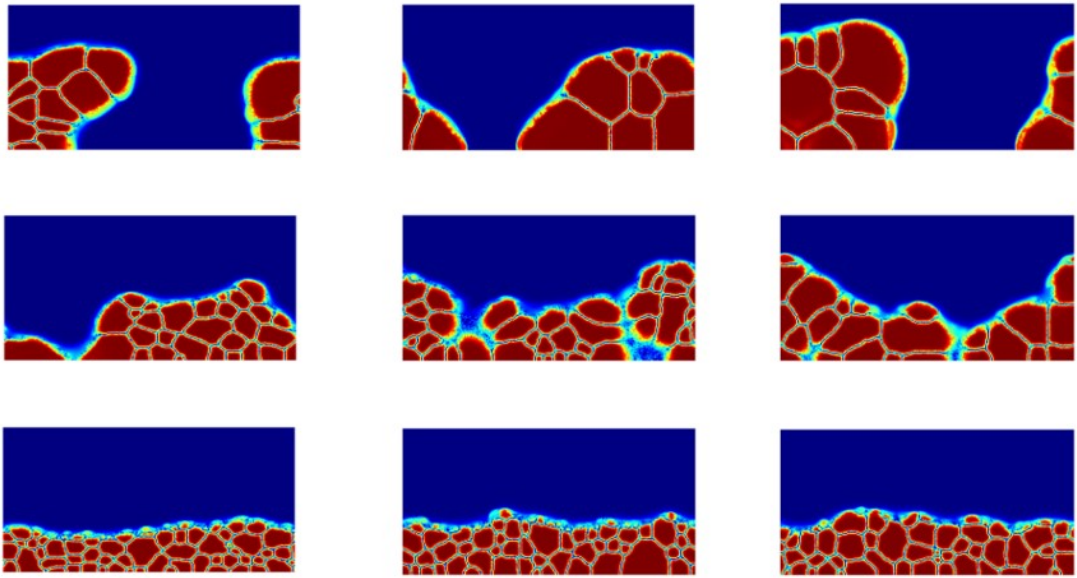

*Figure S 2: Film morphologies at the end of drying. The crystalline order parameter is shown. The different rows correspond to different evaporation rates. From top to bottom:*

$v_{evap} = 67, 201, 536 \ nm \cdot s^{-1}$. *(Left) Viscosities as reported in the table below (Center) Viscosities divided by 10 (Right) Viscosities divided by 100.*

## 1.5. Simulation parameters

| Parameters | Full Name | Value | Unit |
|---|---|---|---|
| **α** | **Evaporation-condensation coefficient** | $(1-9) \cdot 2.3 \cdot 10^{-5}$ | - |
| dx, dy | Grid Spacing | 1 | $nm$ |
| T | Temperature | 300 | K |
| $\rho_i$ | Density | 1000 | $kg/m^3$ |
| $m_1, m_2, m_3$ | Molar Mass | 0.1, 0.1, 0.03 | $kg/mol$ |
| $v_0$ | Molar Volume of the Florry Huggins Lattice Site | $3 \cdot 10^{-5}$ | $m^3/mol$ |
| $\chi_{12,LL}, \chi_{13,LL}, \chi_{23,LL}$ | Liquid – liquid interaction parameter | 0.57, 1, 0 | - |
| $\chi_{12,SL}, \chi_{13,SL}$ | Liquid – solid interaction parameter | 0.15, 0.5 | - |
| $T_m$ | Melting Temperature | 600 | K |
| $L_{fus}$ | Heat of Fusion | 75789 | J/kg |
| W | Energy barrier upon crystallization | 142105 | J/kg |
| $P_0$ | Reference Pressure | $10^5$ | Pa |
| $P_{sat,1}, P_{sat,2}, P_{sat,3}$ | Vapor Pressure | $10^2, 1.5 \cdot 10^3, 10^8$ | Pa |
| $P_{sat,1}, P_{sat,2}, P_{sat,3}$ | Vapor Pressure during annealing | $10^2, 1.5 \cdot 10^2, 10^8$ | Pa |
| $P_i^\infty$ | Partial Vapor Pressure in the Environment | 0 | Pa |
| $E_0$ | Solid-Vapor interaction energy | $5 \cdot 10^9$ | $J/m^3$ |
| $\beta$ | Numerical Free Energy Coefficient | $10^{-5}$ | $J/m^3$ |
| $\kappa_1, \kappa_2, \kappa_3$ | Surface Tension Parameters for Volume Fraction Gradients | $2 \cdot 10^{-10}, 2 \cdot 10^{-10}, 6 \cdot 10^{-9}$ | J/m |
| $\epsilon_c, \epsilon_{vap}$ | Surface Tension Parameters for Order Parameter Gradients | $1.5 \cdot 10^{-5}, 2 \cdot 10^{-4}$ | $(J/m)^{0.5}$ |
| $\epsilon_g$ | Surface Tension parameters for the grain boundaries | 0.2 | $J/m^2$ |
| $D_{s,i}^{\phi_j \rightarrow 1}$ | Self-Diffusion Coefficients in pure materials (all) | $10^{-9}$ | $m^2/s$ |
| $M_c, M_v$ | Allen Cahn mobility coefficients | $4, 10^6$ | $s^{-1}$ |
| $\eta_1, \eta_2, \eta_3$ | Material viscosities | $5 \cdot 10^6, 5 \cdot 10^3, 5 \cdot 10^{-2}$ | Pa·s |
| $D_1^{vap}, D_2^{vap}, D_3^{vap}$ | Diffusion Coefficients in the Vapor Phase | $10^{-16}, 10^{-10}, 10^{-10}$ | $m^2/s$ |
| $t_\phi, t_\varphi, t_{\phi v}$ | Thresholds for crystal detection | $0.4, 0.02, 5 \cdot 10^{-2}$ | - |
| $d_{sl}, c_{sl}, w_{sl}$ | Amplitude, center and with of the penalty function for the diffusion coefficients upon liquid solid transition | $10^{-9}, 0.7, 10$ | - |
| $d_\xi, c_\xi, w_\xi$ | Amplitude, center and with of the penalty function for the order parameter fluctuations | $10^{-2}, 0.85, 15$ | - |
| $d_\eta, c_\eta, w_\eta$ | Amplitude, center and with of the penalty function for the viscosities | $10^{-7}, 0.2, 20$ | - |
| $d_{sv}, c_{sv}, w_{sv}$ | Amplitude, center and with of the penalty function for the Allen Cahn mobility and the solid- vapor interaction energy | $10^{-9}, 0.3, 15$ | - |

## 2. SI to main text section 'Simulation procedure and experimental approach'

The simulations are set up as follows: a 2D cross-section of the film is simulated on 256 x 256 lattice points. Initially, the fluid film is assumed to be fully amorphous and perfectly mixed, and a thin layer of air/vapor phase is placed at the top of the simulation box. The condensed phase is initialized with 20% volume fraction of solute and 80% of solvent (this corresponds roughly to 1.3 M MAPbI$_3$). The initial volume fraction is chosen such that it is well below the crystallization threshold of 26% (see section 0). Periodic boundary conditions are applied in the horizontal direction, Neumann boundary conditions with no flow at the bottom (substrate), and the outflow condition for the evaporating solvent (see equation S19) at the top (vapor phase).

The interaction parameters $\chi_{ij}$ are chosen such that the solute and solvent are completely miscible in the fluid state, but not in the crystalline phase so that the simulated crystals are nearly solvent-free (see section 2.1). Moreover, the resulting equilibrium concentration (saturation concentration $\varphi_s$) of solute in the liquid phase is very low (3.7%). Nucleation and growth are balanced so that the system is neither purely growth nor purely nucleation-dominated. Diffusive constants are chosen so that neither the growth of the crystals nor the evaporation are limited by diffusion, and the amount of solute thus remains homogenous in the liquid phase. The full list of parameters can be found in section 1.5.

Two sets of simulations are performed. The first set solely differs in the evaporation rate of the solvent. The second set only differs in the crystallization rate. The evaporation rate is modified by adjusting the evaporation-condensation coefficient $\alpha$ (see equation S19). The crystallization rate is modified by adjusting the Allen-Cahn mobility *M* (see equation S17). Evaporation-condensation coefficient and Allen-Cahn mobility of the crystalline phase are chosen such that the change in the evaporation rate is sufficient to cover the whole range of possible morphology formation pathways: from crystallization being much faster than evaporation, to crystallization being much slower than evaporation. The range of evaporation rates investigated is nearly one decade and five simulations are performed for each evaporation rate. After a sufficiently long time, when all the untrapped solvent is evaporated, the driving force for evaporation is increased. This is done by increasing the vapor pressure by a factor of ten. This mimics the effect of the annealing step in the experiment. This is a very simplified picture of the annealing process. To describe a quantitatively realistic behavior during this processing step, the temperature-dependence of most of the thermodynamic parameters of the model would has to be taken into account. This is beyond the scope of the present work. Nevertheless, our simplified approach allows to qualitatively reproduce two major evolutions happening during annealing, namely the removal of the remaining solvent and grain coarsening.

There may be solvent remaining in the final stage of the drying because solvent may be trapped either below the crystals or in small channels in between. In such cases, the solvent surface tension energetically hinders further evaporation. Part of the remaining solvent might be removed upon an increase in the solvent vapor pressure, which mimics a thermal annealing process.

## 2.1. Phase diagram of the simulated solute solvent blend and crystal formation mechanisms

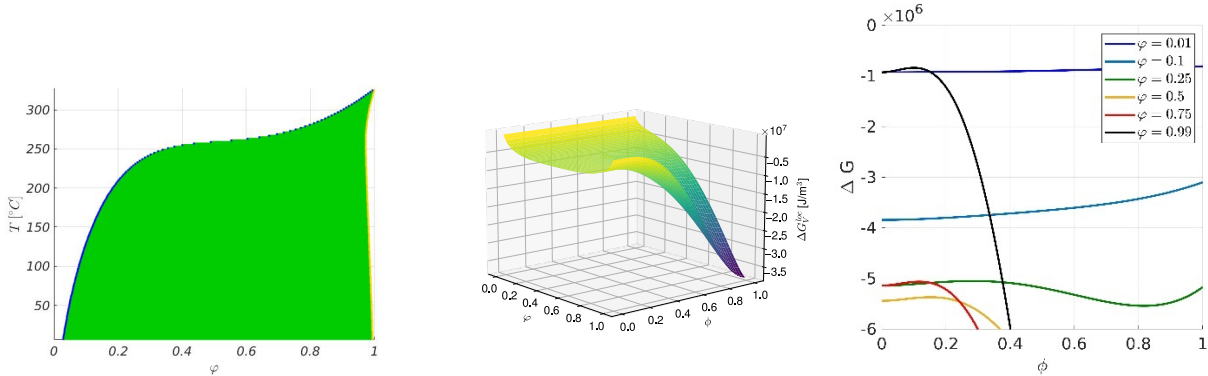

Figure S 3: Left: phase diagram of the investigated system. Blue: liquidus (equilibrium volume fraction of solute in the liquid corresponding to the saturation concentration, $\varphi_s = 0.0374$ at T = 28 °C), yellow: solidus (equilibrium volume fraction of solute in the solid), green: metastable two-phase region. The temperature in the drying simulations is set to 300K (approx. 28 °C). Center: free energy landscape at 300K depending on crystalline order parameter and volume fraction. Note that there is no unstable region, thus crystallization is only possible through nucleation and growth. Right: Free energy plotted against $\phi$ at fixed chosen volume fractions, featuring an energy barrier for $\varphi = 0.25$ to $\varphi = 0.99$. At low volume fractions, the free energy increases with $\phi$, no crystal formation is possible.

A closer look at the free energy landscape helps understanding the crystal formation mechanisms. The figure below shows, beyond the isolines of the free energy surface, (a) the liquidus and solidus points in the 2D space, (b) the location of the maxima (energy barrier) and minima (crystalline phase) along the order parameter variable *at fixed composition*, (c) 'pseudo' spinodal and binodal curves describing the demixing properties {*at fixed order parameter*, and (d) pathways observed during crystal build-up in phase field simulations. There is obviously a 'pseudo' metastable and a 'pseudo' unstable region at high order parameter and intermediate volume fractions. This means that in principle, demixing by nucleation and growth or even spinodal decomposition can take place in a space domain reaching this {composition-crystallinity} region. Now, the system is initially fully amorphous $\phi = 0$), and the amorphous phase is stable whatever the composition. Only the fluctuations of the order parameter allow the energy barrier for crystallization to be overcome, on some limited domains in space. Then, these domains, which are actually the crystals nuclei, quickly evolve towards higher order parameter and volume fraction values (stable solidus point) while the amorphous phase evolves towards the stable liquidus point. These are precisely the features of a nucleation process. Note that nucleation is only possible for sufficiently large solute volume fractions ($\varphi > 0.2$). Thereby, for the highest solute concentrations, the crystallization pathways mostly avoid the 'pseudo' unstable and metastable regions. For lower concentrations however, the crystallization pathway may go through it, and one might therefore expect demixing to take place inside the nuclei. But in practice this does not happen. The reason is the following: the unstable and metastable regions of the free energy landscape are just a necessary condition, but demixing requires some space[10] and time[11] [12] to take place. Indeed, the nuclei are too small and their composition/crystallinity evolves too quickly, which prevents demixing inside them. Instead, the composition in the nuclei progresses quickly towards the 'pseudo'-binodal composition (see the example of the nuclei in the $\varphi = 0.25$ mixture).

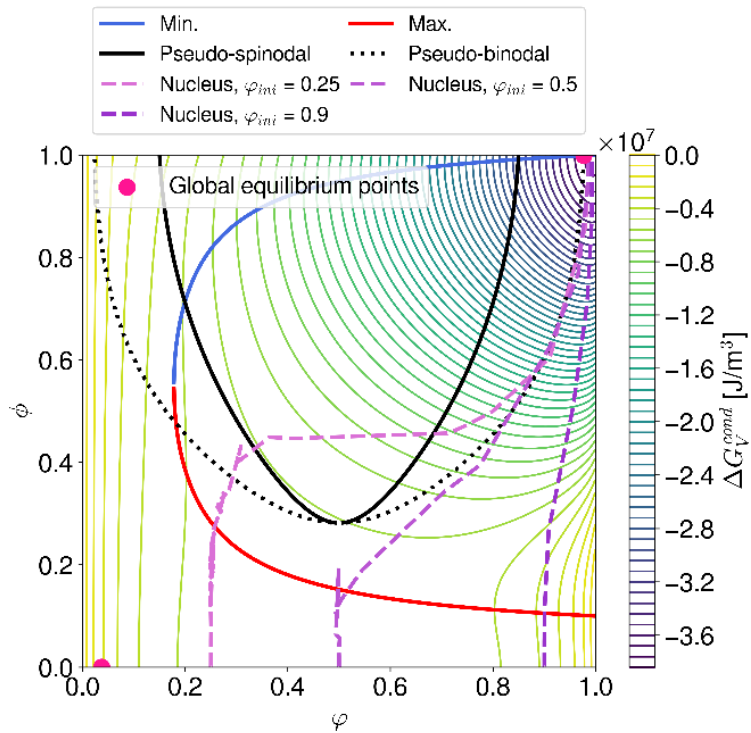

*Figure S 4: 2D contour plot of the free energy surface, featuring the liquidus and solidus points in the 2D space (pink points), the location of the energy maxima (red curve) and minima (blue curve) along $\phi$ at fixed $\varphi$, the 'pseudo' spinodal (black full line) and binodal (black dotted line). The typical trajectories observed during crystal build-up, starting from a volume fraction of solute $\varphi = 0.25$, $\varphi = 0.5$ and $\varphi = 0.9$, respectively, are shown by the dashed violet lines.*

The series of snapshops below show the evolution of the volume fraction and parameter field during nucleation in a binary mixture with 25%, 50% and 90% of solute, corresponding to the three trajectories reported on the free energy surface above. Thereby, note that the local volume fraction variations are the consequences of local order parameter variations overcoming the energy barrier.

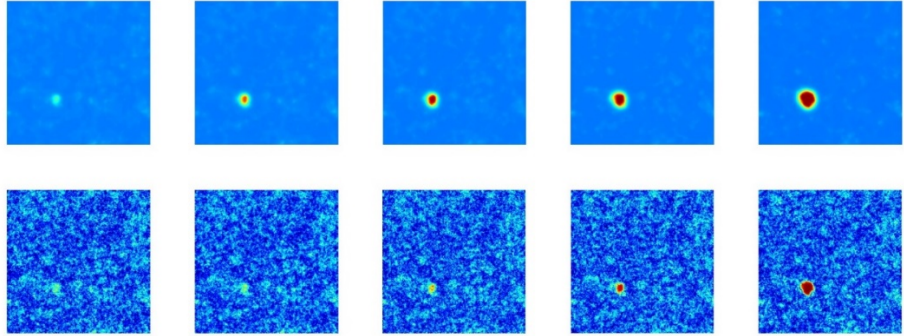

*Figure S 5: Volume fraction field (top row) and order parameter field (bottom row) at increasing times (from left to right) during crystallization in a binary blend, $\varphi = 0.25$ of solute.*

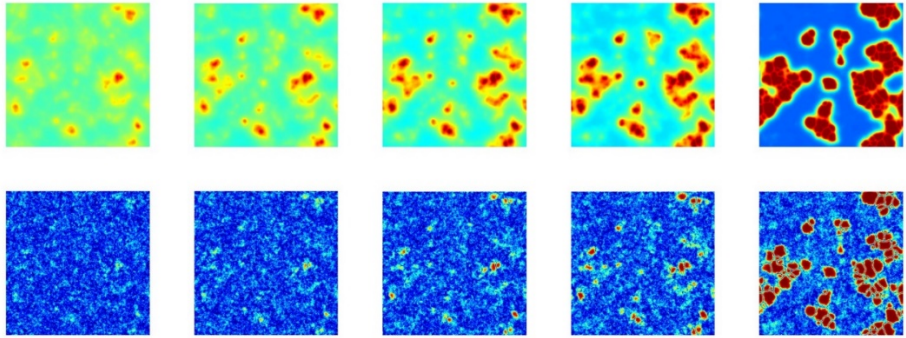

*Figure S 6: Volume fraction field (top row) and order parameter field (bottom row) at increasing times (from left to right) during crystallization in a binary blend, $\varphi = 0.5$ of solute.*

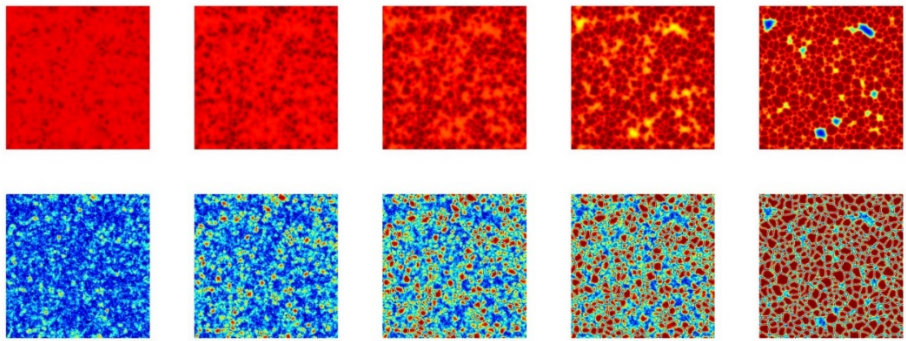

*Figure S 7: Volume fraction field (top row) and order parameter field (bottom row) at increasing times (from left to right) during crystallization in a binary blend, $\varphi = 0.9$ of solute.*

As a conclusion, with the thermodynamic parameters used in this work, crystal form exclusively by one step nucleation and growth. The reader is referred to previous work for a more general discussion on other possible nucleation pathways with different thermodynamic properties.[13]

## 2.2. Critical volume fraction (onset of nucleation)

In these simulations the overall composition is kept constant (the solvent may not evaporate). The simulation is run until the solute is fully crystallized. The time $t_{1/2}$ until half of the solute is crystallized is extracted from the data. Two-dimensional simulations containing solute and solvent are performed.

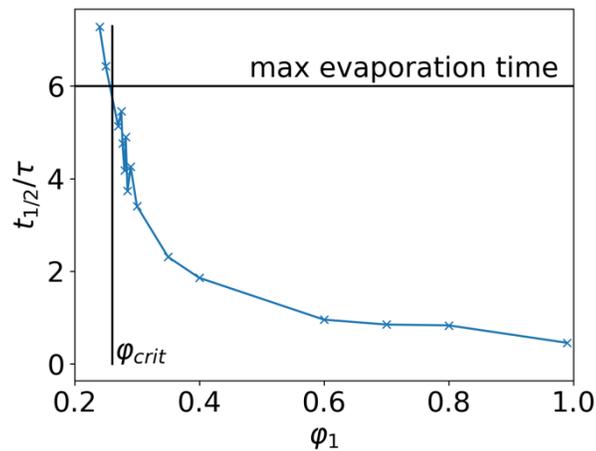

Figure S 8: Crystallization half time $t_{1/2}$ measured in a binary blend. The time increases for decreasing volume fractions. Below some (low) volume fraction there is not enough material, the driving force for crystallization is too low and the crystallization time diverges. The quantity of interest $\varphi_{crit}$ is the critical volume fraction for which crystallization cannot occur within the evaporation time of the drying simulations. The time for evaporation is maximally 6 $\tau$. The intersection of the crystallization half-time with the maximal time of evaporation is $\varphi_{crit}$, which is roughly 0.26.

## 2.3. Time series of a single simulation with a low evaporation rate

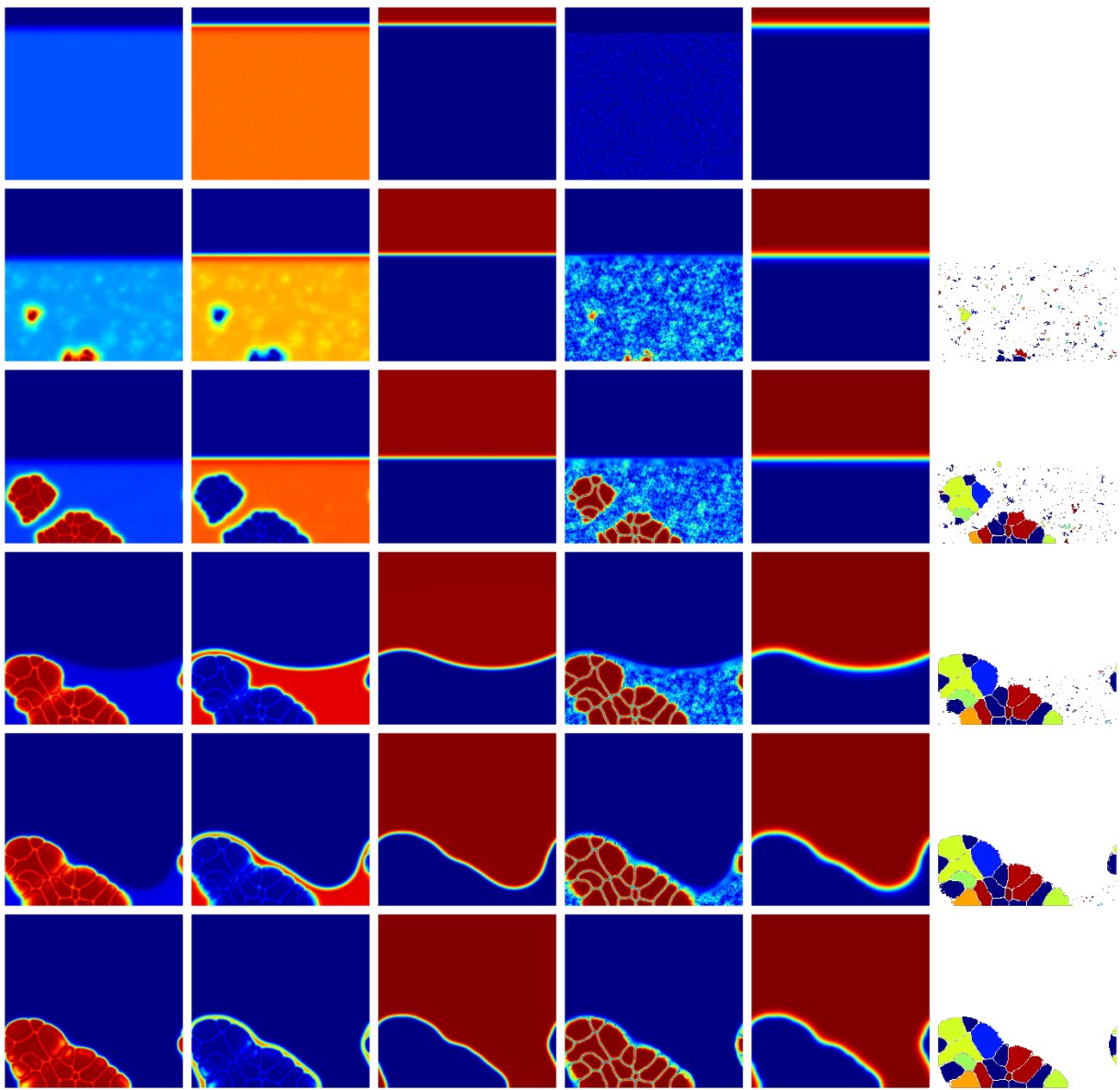

*Figure S 9: Time series for one single simulation with a low drying rate. The time increases with each row from top to bottom. From left to right: solute volume fraction ($\varphi_1$), solvent volume fraction ($\varphi_2$), air volume fraction ($\varphi_3$), crystalline order parameter ($\phi_c$), vapor order parameter ($\phi_{vap}$) and the orientation parameter ($\theta$).*

## 2.4. Experimental methods

*Materials*:

Lead iodide (PbI2, 99%), methylammonium iodide (MAI, 98%), Benzyl Chloride (CB, 99%) and Poly-(3-hexylthiophen-2,5-diyl) (P3HT) were purchased from Sigma. Anhydrous 2-Methoxyethanol (2ME, Aldrich, 99.8%) and 1-methyl-2-pyrrolidinone (NMP, 99.8%) were purchased from Aldrich. Tin (IV) oxide ($SnO_2$, 15% in H2O colloidal dispersion) was purchased from Alfa Aesar). Carbon paste was purchased from Liaoning Huite Photoelectric Technology Co. Ltd. All the chemicals were used as received without further purification.

*Gas-quenching-assisted blade deposition of perovskite films:*

Equal molar amounts of MAI and $PbI_2$ were dissolved in anhydrous 2ME and NMP, (2ME:NMP, v:v=37:3) to prepare 1 M $MAPbI_3$ stock solution and stirred at room temperature for 1 h. 20 µL of the precursor solution was doctor-bladed onto a 25 mm x 25 mm glass substrate at 3 mm s$^{-1}$ and a gap height of 150 µm. After casting, the wet film was blown from the top with a continuous flow of dry air for 60 s, which is denoted as "gas quenching". The air pressure can be controlled from 0 Bar to 5.0 Bar. Following that, the films were thermally annealed at 100°C by a heat gun for 10 minutes. Blade coating of the perovskite precursor films was carried out on a commercial blade coater (ZAA2300.H from ZEHNTNER) using a ZUA 2000.100 blade (from ZEHNTNER) at room temperature in air.

*Solar cell fabrication*:

The pre-patterned indium tin oxide (ITO) coated glass (Liaoning Huite Photoelectric Tech. Co., Ltd.) was sequentially cleaned by sonicating the substrates in acetone and isopropanol for 15 min each. Then, the substrates were treated in an UV–Ozone box for 20 min to remove organic residues and to enable better wetting. An aqueous $SnO_2$ nanoparticle solution was used to prepare the electron transport layer. The solution was diluted to 5.0 wt% $SnO_2$ and treated in the ultrasonic bath for 10 min before filtering with a 0.45 µm PTFE filter. The solution was then doctor-bladed at 80 °C and 15 mm s$^{-1}$, and a gap height of 100 µm. Subsequently, the film was annealed at 150 °C for 30 min to form a compact layer. The perovskite absorber layer was subsequently deposited using the gas-quenching-assisted blade-coating method described above.

For the hole transport layer, 10 mg mL$^{-1}$ P3HT was dissolved in anhydrous CB and stirred at 80 °C for at least 2 h. A gap height of 150 µm and a volume of 40 µL was used for doctor-blading P3HT solutions. The coating temperatures and speeds for P3HT were 60 °C and 5 mm s$^{-1}$, respectively. After coating the P3HT layer, the film was annealed at 100 °C for 5 min. For the carbon electrode, the carbon paste was stencil-printed on the as-prepared film and annealed at 120 °C for 15 min. For this, the electrode pattern was cut out of an adhesive tape with a laser. The tape was then placed on the substrate with the sticky side down. The cutouts were filled with carbon paste by blade coating. The tape was then removed carefully and the substrate was annealed on a hot plate at 100 °C for 15 min.

*Characterizations:*

Solar cells were characterized by measuring their current–voltage (J–V)-characteristics with an AAA solar simulator, which provides AM1.5G illumination and source measurement system from LOT-Quantum Design, calibrated with a certified silicon solar module. The voltage sweep range was −0.5 to 1.5 V in steps of 20 mV. Morphologies of the perovskite films were imaged with a confocal microscope (FEI Apreo LoVac).

Scanning electron microscopy (SEM): A FEI Helios Nanolab 660 was used to acquire SEM images and to prepare FIB cross-sections. The final polishing with the ion beam was performed at 5 kV and 80 pA.

X-ray powder diffraction (XRD): X-ray diffraction analysis was performed by classical ex-situ Bragg–Brentano geometry using a Panalytical X'pert powder diffractometer with filtered Cu-Kα radiation and an X'Celerator solid-state stripe detector.

Transmittance and reflectance spectra of the samples were carried out using a UV-VIS-NIR spectrometer (Lambda 950, from Perkin Elmer). For the haze measurement, the diffuse transmittance and total transmittance were detected without or with a reflection standard placed, respectively. The detector with R955 PMT works at the wavelength of 160 nm to 900 nm.

The roughness and thickness of the perovskite films were measured by confocal microscope µsurf custom from NanoFocus AG.

In situ white light reflectance spectrometer (WLRS, Thetametrisis): For high-quality reflective measurements, all the film was deposited on the silicon wafers which were cut into 1 × 1 cm substrates. The refractive index (n) and extinction coefficient (k) of the perovskite wet films were set as 1.5 ± 0.5 and 0.3 ± 0.1, respectively.

In Situ PL: PL measurements were acquired on a home-built confocal setup using a 532 nm or 450 nm laser diode, a plano-convex lens above the substrate, a 550 nm long-pass filter, and a fiber-coupled spectrometer (AVANTES, ULS2048XL Sensline series) calibrated by the manufacturer. The distance between the plano-convex lens and the substrate was optimized such that the PL intensity of a dry film was maximized. The working distance was not adjusted with the change of the wet film thickness during the drying process.

In situ UV-vis: The in-situ absorption measurements were performed using a F20-UVX spectrometer (Filmetrics, Inc.) equipped with tungsten halogen and deuterium light sources (Filmetrics, Inc.). The signal is detected with the same fiber-coupled spectrometer with a spectral range of 300 to 1000 nm. Most of the measurements were performed with an integration time of 0.5 s (thin perovskite layer) per spectrum. The UV–vis absorption spectra are calculated from the transmission spectra, using the following equation: $A\lambda = -\log_{10}(T)$, where $A\lambda$ is the absorbance at a certain wavelength ($\lambda$) and T is the calibrated transmitted radiation.

# 3. SI to main text section 'Impact of the drying rate on the morphology and model validation'

## 3.1. Infrared reflectometry and XRD spectra

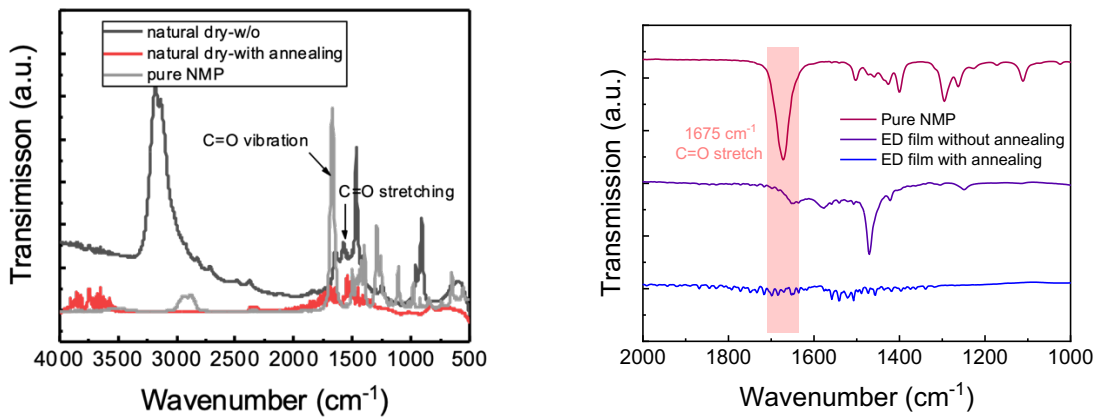

*Figure S 10: Infrared reflectometry spectra of the environmentally dried film, the film after annealing, and for pure NMP.*

The intense peak corresponds to C=O symmetric stretching at 1675 cm-1 for pure NMP and a weak peak at similar position for environmentally dry film, while an absence of this peak in the environmentally dry film after thermal annealing treatment. Therefore, we confirmed that the NMP could left in the film if no gas quenching treatment is performed, which is in good agreement with the XRD patterns.

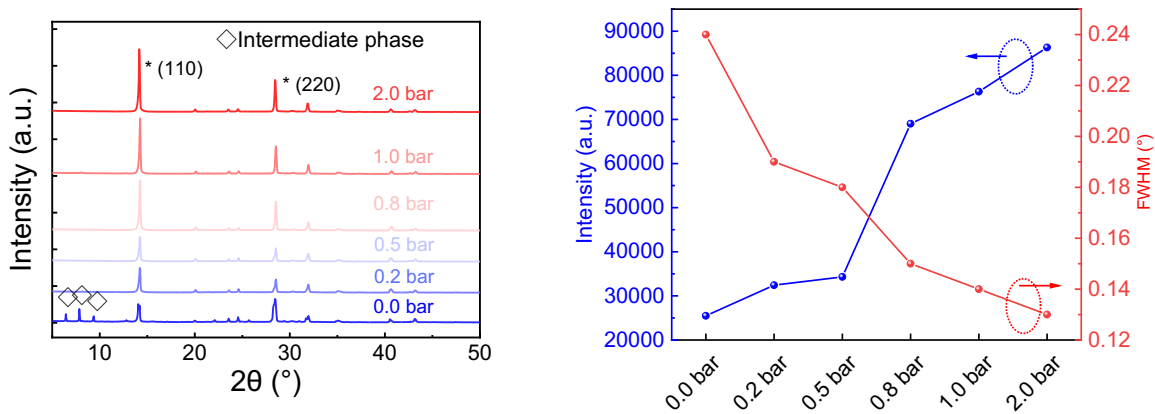

*Figure S 11: Left: XRD spectra for all the experimental drying conditions. Right: full-width at half maximum (FWHM) and intensity of the (110) peak of all the samples.*

Regarding the XRD patterns of ED film, we found the peak located at 6.4°, 7.8°, and 9.3°, which indicates the lattice of the PbI2 crystal has been enlarged by large molecules. The large molecules could be the NMP because some reports mentioned the PbI2(NMP) XRD peak located at 8.1°. The peak position is determined by the crystal lattice, and layered PbI2 have weak bonding, which allows the insertion of different guest molecules by van der Walls interactions.

As shown in Figure S11, a set of preferred orientations at 14.15°, 28.44°, 31.85°, 40.62° and 43.14° was observed, with these assigned to the (110), (220), (310), (224) and (330) planes of the MAPbI3 perovskite tetragonal structure, respectively. Minor peaks of the (200), (211), and (202) planes are present at 2θ values of 20.03°, 23.50°, and 24.55°, respectively, clearly indicating that all perovskite films are of high phase purity.

If necessary, we can plot the FWHM and intensity of XRD peak at 14.15°. The FWHM and intensity of the (110) peak are shown in Figure S11b, the crystallinity of the perovskite films was increased by increasing the evaporation rate.

## 3.2. Visualization of trapping mechanisms of solvent in the film after drying

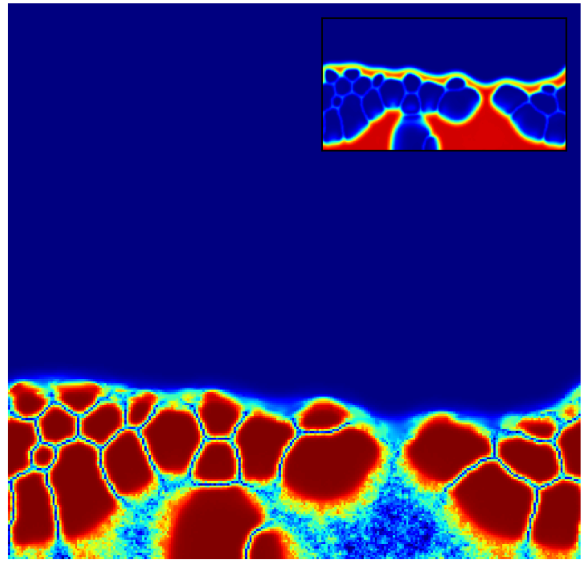

*Figure S 12: Exemplary dry state of a simulation with slow to medium evaporation rate. The crystalline order parameter and the solvent volume fraction are displayed (as inset). The solvent on the left side is completely trapped between the crystals and the substrate and can therefore hardly evaporate. The solvent trapped on the right side cannot evaporate due to surface tension effects: further evaporation would require a tremendous increase of the liquid meniscus curvature, which is associated with an unaffordable surface energy increase.*

## 3.3. Evolution of the film height and crystallinity during annealing

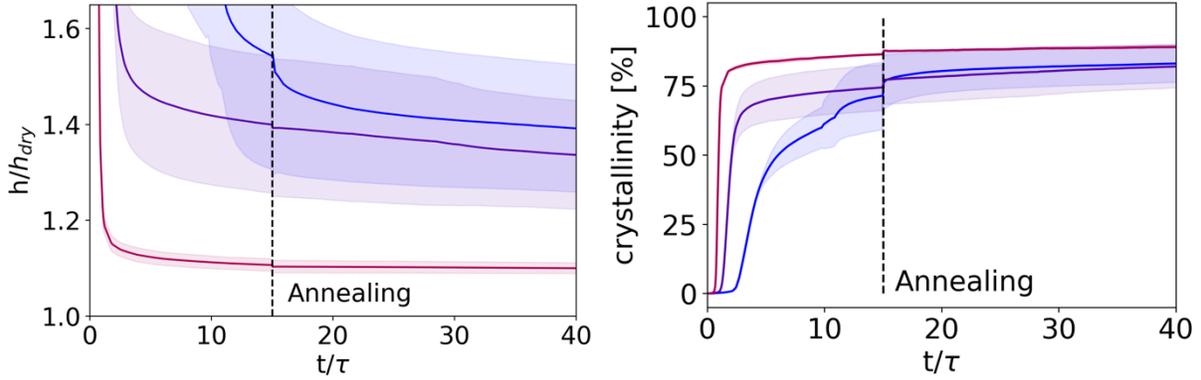

*Figure S 13: Left: Simulated Film Height including annealing. A slight drop in film height is visible at the start of annealing, which has two reasons. First, the evaporation rate is enhanced dramatically because the vapor pressure is increased abruptly. Therefore, previously trapped solvent now evaporates quickly. Second, the film-vapor interface is modified, which is a purely numerical effect. Unfortunately, the magnitude of these two effects could not be evaluated independently. Right: Crystallinity of the system, including annealing. The increase of crystallinity at the start of annealing is a numerical artefact also due to the modification of the diffuse film-vapor interface structure at the onset of annealing. This has a noticeable but limited impact on the calculation of the overall crystallinity inside the film.*

## 3.4. Evaluation of crystal size, amount of uncovered substrate, and roughness in the simulation

The fraction of uncovered substrate is the fraction of vertical lines in the simulation, where the volume fraction of solute has no value larger than 0.8. The volume fraction of the solute is chosen instead of the crystalline order parameter for this coverage evaluation, because the crystalline order parameter field is noisy due to the applied fluctuations. The value of 0.8 is chosen because beyond $\phi_c = 0.8$, the solute is always in the crystallized state in the dry film.

For the roughness calculation, the highest point $h_i$ in each vertical line of the simulation box, surpassing a solute volume fraction of 0.8 is used as an upper boundary of the film. The roughness $R_Q$ is then calculated as

$$R_Q = \sqrt{\frac{\sum_{i=1}^{N}(h_i - h_{final})^2}{N}} \qquad (S26)$$

where *N* is the number of columns.

For the crystal sizes, the equivalent radius of each individual grain $r_i$ (defined as the domains with homogenous/identical orientation value) is calculated. The average crystal sizes are calculated as:

$$r = \frac{\sum_{i=1}^{N} v_i r_i}{V} \tag{S27}$$

where *V* is the total volume of the crystals and $v_i$ the fraction the volume of crystal *i*.

## 3.5. SEM images

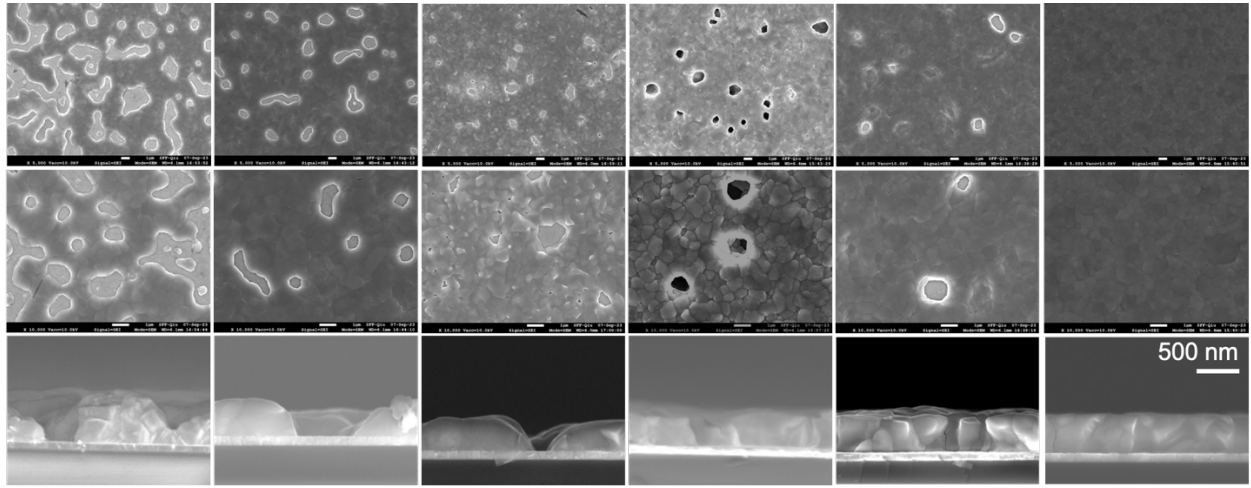

*Figure S 14: First and second row: SEM top views at different magnifications after annealing. Third row: SEM cross-section. From left to right: 0, 0.2, 0.5, 0.8, 1, 1.5, and 2 bar air pressure for gas quenching during fabrication.*

## 3.6. Simulated film morphologies after drying and annealing

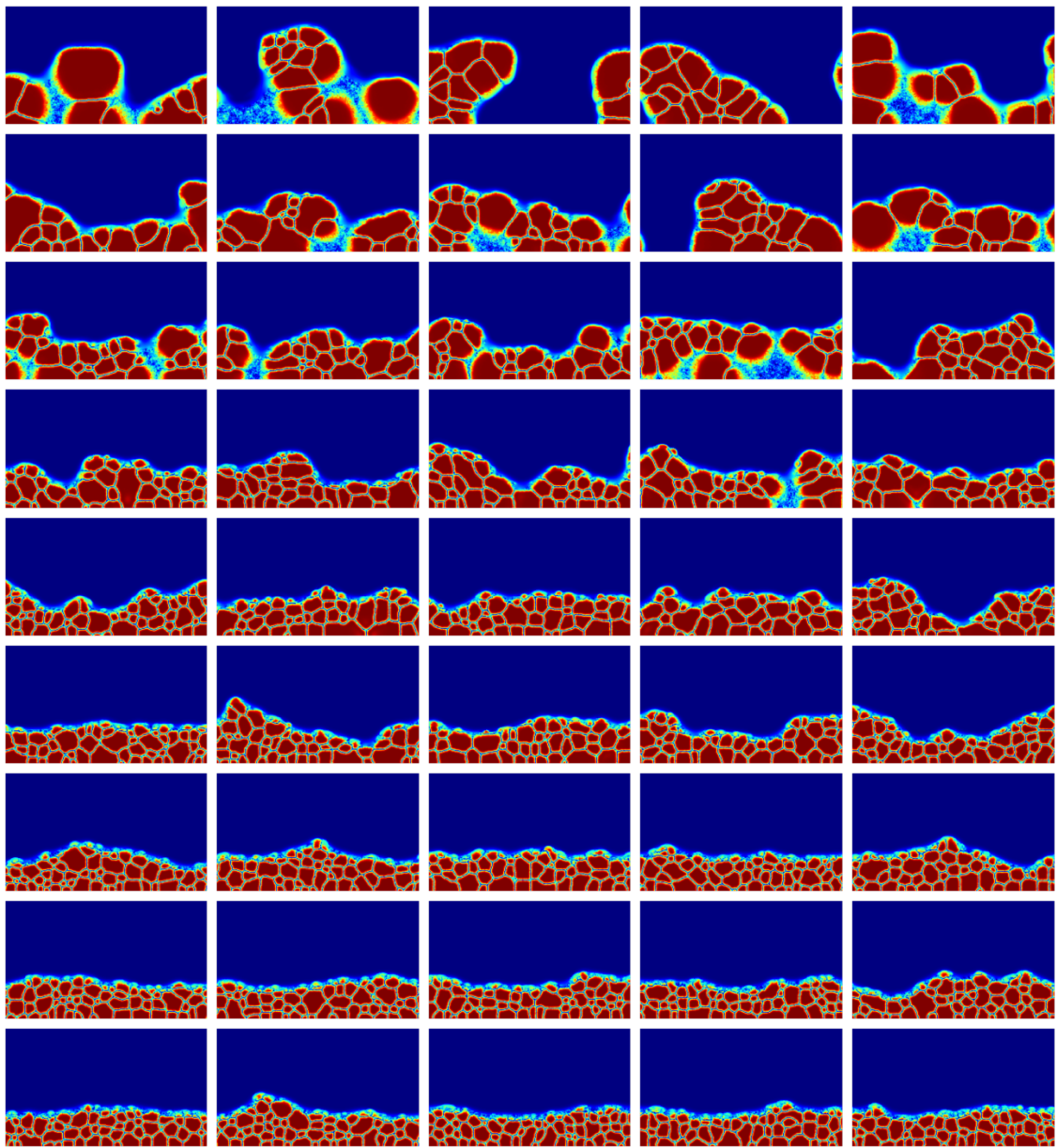

*Figure S 15: Film morphologies at the end of drying. The crystalline order parameter is shown. The different rows correspond to different evaporation rates. From top to bottom: $v_{evap}$ = 67, 134, 201, 268, 335, 402, 469, 536, 603 nm/s. The columns represent five different runs with exactly the same simulation parameters, including $v_{evap}$.*

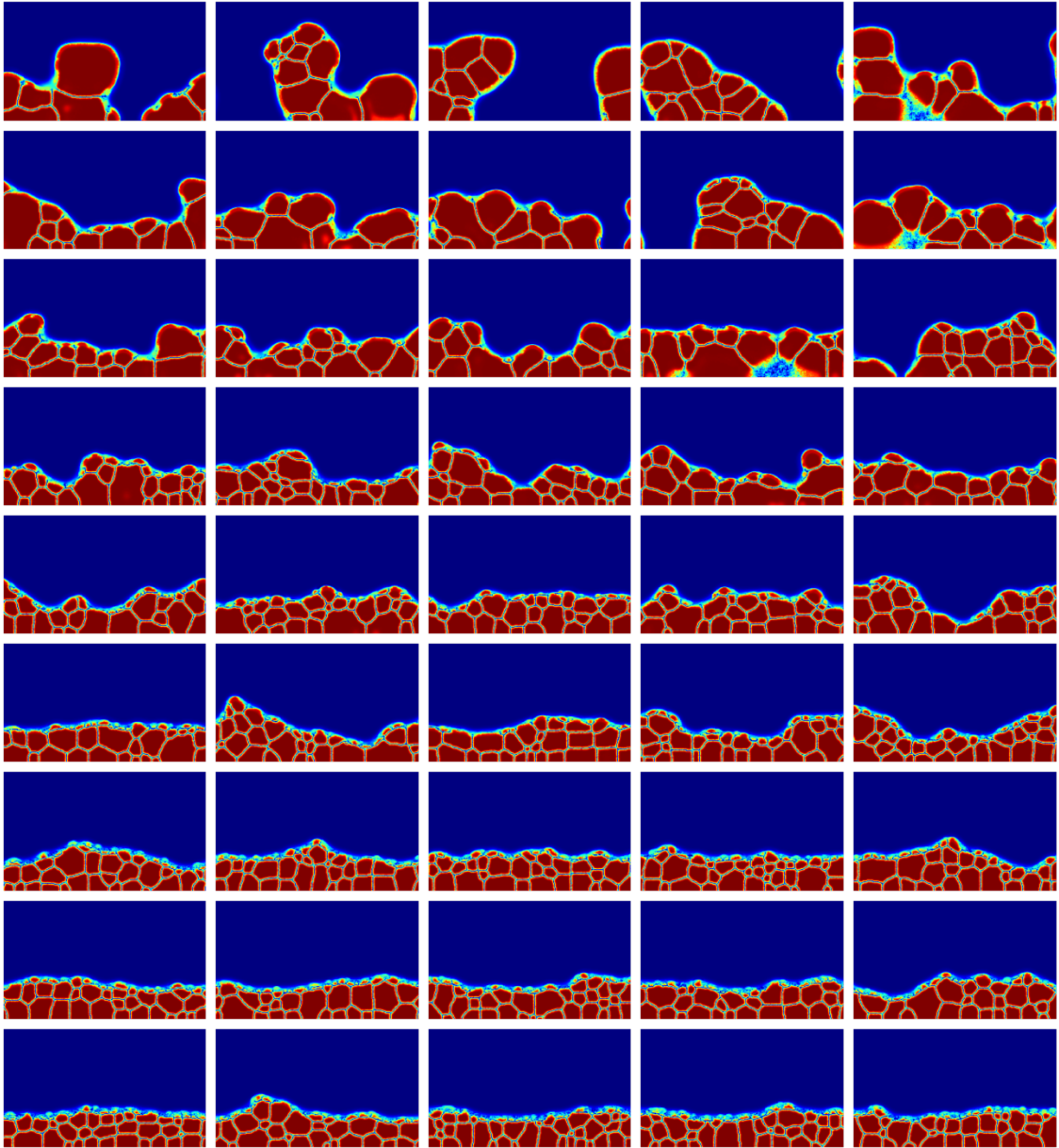

*Figure S 16: Film morphology at the end of annealing (30s timespan). The crystalline order parameter is shown. The different rows correspond to different evaporation rates. From top to bottom: $v_{evap}$ = 67, 134, 201, 268, 335, 402, 469, 536, 603 nm/s. The columns represent five different runs with exactly the same simulation parameters, including $v_{evap}$.*

## 3.7. Comparison between the film morphology after drying and after annealing

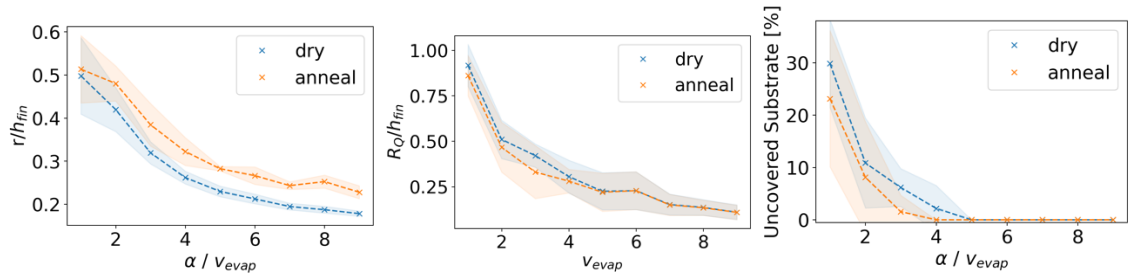

*Figure S 17: Comparison between morphology descriptors after drying and after 30s annealing. (Left) the crystal sizes increase during annealing due to grain coarsening. For low evaporation rates, the morphological features only a few, large and separated crystals and this effect is less pronounced. (Center) Film roughness: The roughness of the film stays the same or decreases slightly, if crystals at the solid-vapor interface disappear due to coarsening. (Right) Uncovered substrate: The uncovered substrate decreases during annealing due to coarsening.*

## 3.8. Haze factor

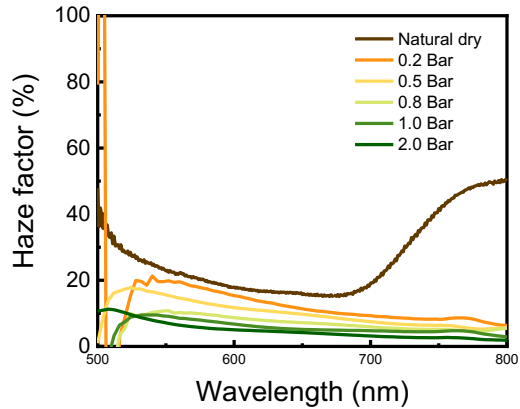

*Figure S 18: Haze factor: The haze factor is a measurement to evaluate the light-scattering ability of the thin film. The haze factor is calculated from the ratio of diffuse and total transmission. The natural dry film shows a very high haze at the long wavelength which could be attributed to the uncovered area, resulting in the high light scattering. For the films processed with a higher air flow rate, i.e., 1.0 bar, the haze factor is lower than 10%, indicating that the film has a smooth surface, such that the light is absorbed with low scattering.*

## 3.9. PL grain size and UV-vis spectra

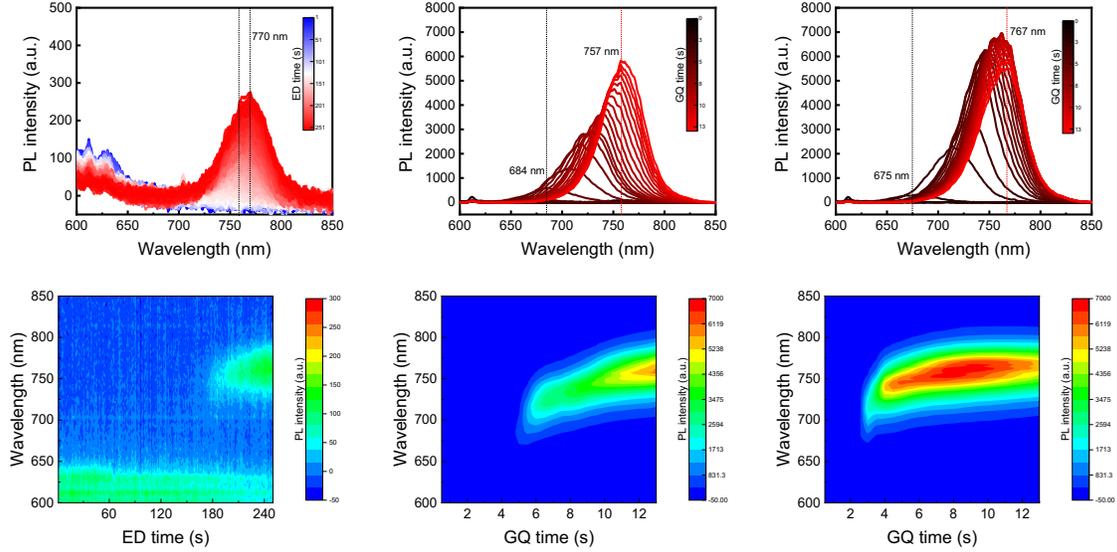

*Figure S 19: Measured PL curves for environmental drying, 0.2 bar, and 2 bar (left to right). First row: Spectra, second row: heat map.*

The grain sizes of the crystals can be calculated with [14,15]

$$E_g = E_{g,bulk} + \frac{2\pi^2 \hbar^2}{m_e d^2},  \tag{S28}$$

where $E_{g,bulk}$ is the energy gap of the bulk material, $\hbar$ is the reduced Planck constant, $m_e$ is the effective mass of the excitons, and $d$ the average size of the grains.

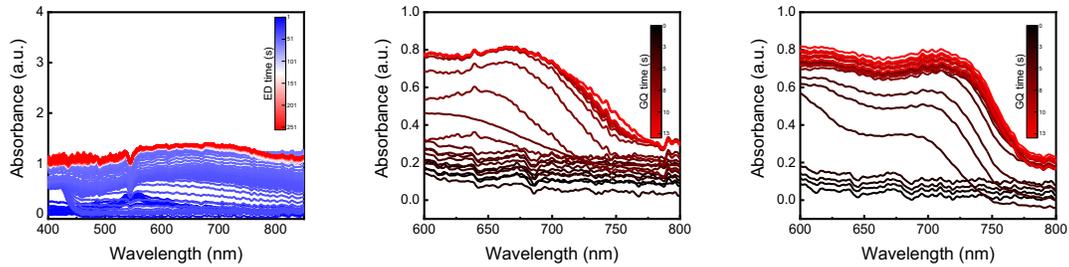

*Figure S 20: Measured UV-vis for environmental drying, 0.2 bar, and 2 bar air pressure (left to right).*

# 4. SI to main text section 'Dependence of the device performance on film morphology'

## 4.1. Stabilized power output

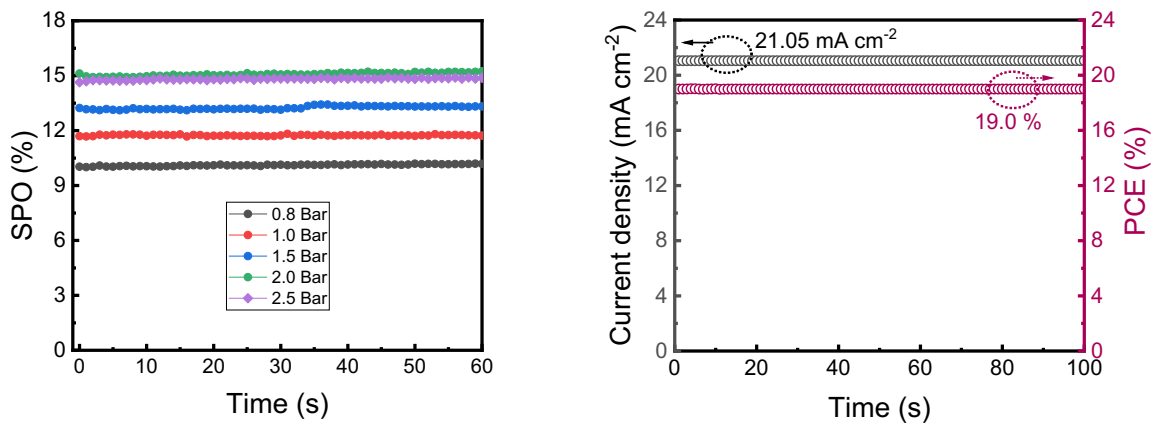

*Figure S 21: Left: Stabilized power output. Right: time-dependent photocurrent and PCE at the maximum power point of the champion cell.*

## 4.2. JV-scan of the champion cell and device yield

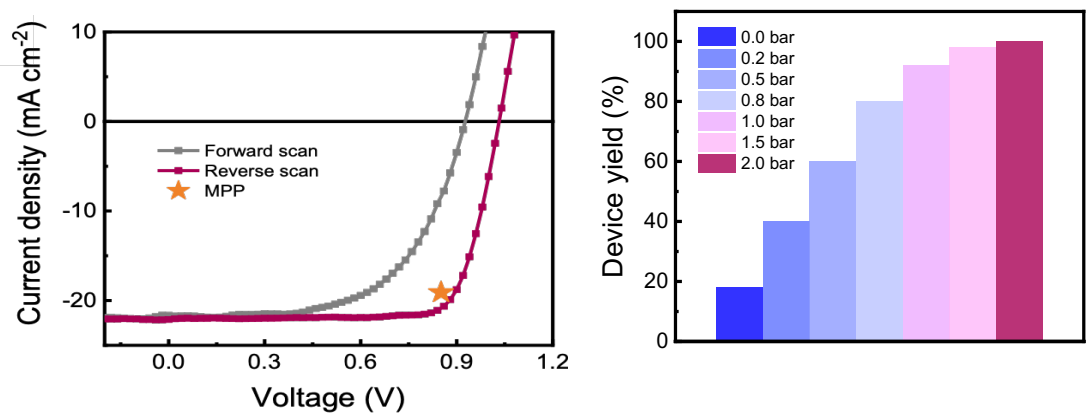

*Figure S 22: Left: Forward and Reverse scan of the champion cell. Right: Device yield of the measured solar cells depending on the perovskite films prepared by various evaporation rates (0, 0.2, 0.5, 0.8, 1.0, 1.5, and 2.0 bar, respectively).*

## 4.3. Drift-diffusion simulations

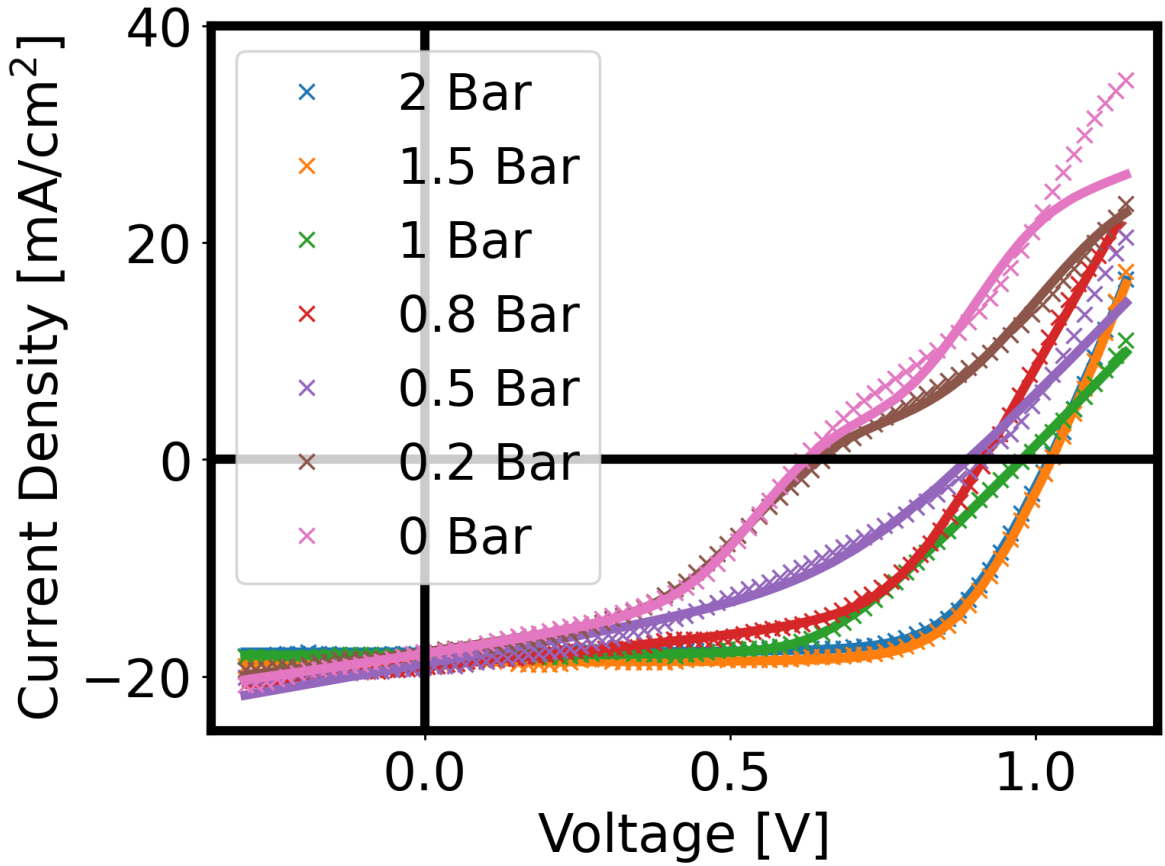

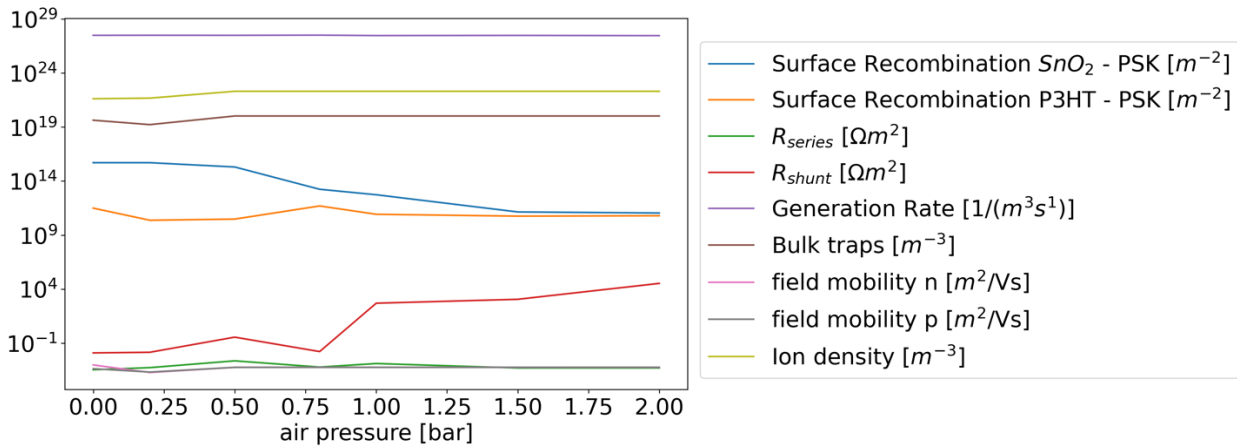

*Figure S 23: Drift diffusion simulations. First Row: experimental (crosses) and simulated (full line) JV curves. Fitted are the trap densities at the interfaces, the generation rate, the series, and the shunt resistance. The electron and hole mobilities, the trap density in the bulk, as well as the ion density, are kept constant, except for 0 bar and 0.2 bar vapor pressure. No reasonable fit could be achieved for these two pressures while keeping these three parameters constant. The obtained values are displayed in the second row. The shunt resistance decreases for lower air pressures and the trap density increases. The*

*other parameters remain approximately constant. Note also that the weak slope of the JV curve at high voltage and its S-shape cannot be explained by a mere variation of the series resistance.*

| Parameters | Full Name | Value | Unit |
|---|---|---|---|
| T | Temperature | 295 | $K$ |
| L | Total thickness of the device | $470 \cdot 10^{-9}$ | $m$ |
| eps_r | Relative dielectric constant | 24 | |
| CB | Conduction band edge | 3.9 | $eV$ |
| VB | Valence band edge | 5.49 | $eV$ |
| Nc | Effective density of states | $5 \cdot 10^{24}$ | $m^{-3}$ |
| n_0 | Ionised n-doping | 0 | $m^{-3}$ |
| p_0 | Ionised p-doping | 0 | $m^{-3}$ |
| L_TCO | Tickness of the ITO layer | $110 \cdot 10^{-9}$ | $m$ |
| L_BE | Tickness of the back electrode | $200 \cdot 10^{-9}$ | $m$ |
| lambda_min | Minimum wavelength of the spectrum for the calculated generation profile | $350 \cdot 10^{-9}$ | $m$ |
| lambda_max | Maximum wavelength of the spectrum for the calculated generation profile | $800 \cdot 10^{-9}$ | $m$ |
| **mun_0** | **Electron mobility at zero field** | **$6 \cdot 10^{-4}$ (fitted for 0, 0.2 bar)** | $m^2/Vs$ |
| **mup_0** | **Hole mobility at zero field** | **$6 \cdot 10^{-4}$ (fitted for 0, 0.2 bar)** | $m^2/Vs$ |
| mob_n_dep | Electron mobility | 0, constant | |
| mob_p_dep | Hole mobility | 0, constant | |
| W_L | Work function of the left electrode | 4.25 | $eV$ |
| W_R | Work function of the right electrode | 5.1 | $eV$ |
| Sn_L | Surface recombination velocity of electrons at the left electrode | $-1 \cdot 10^{-7}$ | $m/s$ |
| Sp_L | Surface recombination velocity of holes at the left elclectord | $-1 \cdot 10^{-7}$ | $m/s$ |
| Sn_R | Surface recombination velocity of electrons at the right electrode | $-1 \cdot 10^{-7}$ | $m/s$ |
| Sp_R | Surface recombination velocity of holes at the right electrode | $-1 \cdot 10^{-7}$ | $m/s$ |
| **Rshunt** | **Shunt resistance** | **$5 \cdot 10^{3}$ (fitted)** | $\Omega m^2$ |
| **Rseries** | **Resistance place in series with the device** | **$2 \cdot 10^{-4}$ (fitted)** | $\Omega m^2$ |
| L_LTL | Thickness of the left transport layer | $20 \cdot 10^{-9}$ | $m$ |
| L_RTL | Thickness of the right transport layer | $50 \cdot 10^{-9}$ | $m$ |
| Nc_LTL | Effective density of states of the left transport layer | $2.7 \cdot 10^{24}$ | $m^{-3}$ |
| Nc_RTL | Effective density of states of the right transport layer | $5 \cdot 10^{26}$ | $m^{-3}$ |
| doping_LTL | Density of ionized dopants of the left transport layer | 0 | $m^{-3}$ |
| doping_RTL | Density of ionized dopants of the right transport layer | 0 | $m^{-3}$ |
| mob_LTL | Mobility of electrons and holes in the left transport layer | $5 \cdot 10^5$ | $m^2/Vs$ |
| mob_RTL | Mobility of electrons and holes in the right transport layer | $5 \cdot 10^7$ | $m^2/Vs$ |
| nu_int_LTL | Interface transfer velocity between the main layer and the left transport layer | $1 \cdot 10^3$ | m/s |
| nu_int_RTL | Interface transfer velocity between the main layer and the right transport layer | $1 \cdot 10^3$ | m/s |
| eps_r_LTL | Relative dielectric constant of the left transport layer | 10 | |
| eps_r_RTL | Relative dielectric constant of the right transport layer | 3 | |
| CB_LTL | Conduction band edge of the left transport layer | 4.2 | eV |
| CB_RTL | Conduction band edge of the right transport layer | 3 | eV |
| VB_LTL | Valence band edge of the left transport layer | 8.4 | eV |
| VB_RTL | Valence band edge of the right transport layer | 5.15 | eV |
| TLsGen | Transport layer absorption | 0, no | |
| TLsTraps | Transport layer contain traps | 0, no | |
| InosInTLs | Ions can move from the bulk into the transport layers | 0, no | |
| **CNI** | **Concentration of negative ions** | **$2 \cdot 10^{22}$ (fitted for 0, 0.2 bar)** | $m^{-3}$ |
| **CPI** | **Concentration of positive ions** | **$2 \cdot 10^{22}$ (fitted for 0, 0.2 bar)** | $m^{-3}$ |
| mob_ion_spec | Which ionic species can move | 1, only positive | |
| ion_red_rate | Rate at which the ion distribution is updated | 1 | |
| **Gehp** | **Average generation rate of the electron-hole pairs in the absorbing layer** | **$2.83 \cdot 10^{27}$ (fitted)** | $m^{-3}s^{-1}$ |
| Gfrac | Actual average generation rate as a fraction of Gehp | 1 | |
| Gen_profile | File of the generation profile | None, uniform | |
| Field_dep_G | Field-dependent splitting of the electron-hole pairs | 0, no | |
| kdirect | Rate of direct recombination | $1.6 \cdot 10^{-17}$ | $m^3/s$ |
| UseLangevin | Constant rate of recombination of Langevin expression | 0, direct recombination | |
| **Bulk_tr** | **Density of traps in the bulk** | **$1.04 \cdot 10^{20}$ (fitted)** | $m^{-3}$ |
| **St_L** | **Number of traps per area at the left interface between the left transport layer and the main absorber** | **$2 \cdot 10^{12}$ (fitted)** | $m^{-2}$ |
| **St_R** | Number of traps per area at the left interface between the left transport layer and the main absorber | $1 \cdot 10^{10}$ (fitted) | $m^{-2}$ |
| num_GBs | Number of grain boundaries | 0 | |
| GB_tr | Number of traps per area at a grain boundary | $1 \cdot 10^{13}$ | $m^{-2}$ |
| Cn | Capture coefficient for electrons (for all traps) | $1 \cdot 10^{-13}$ | $m^3$ |
| Cp | Capture coefficient for holes (for all traps) | $1 \cdot 10^{-13}$ | $m^3$ |
| ETrapSingle | Energy level of all traps | 4.91 | eV |
| Tr_type_L | Traps at the left interface | -1, acceptor | |
| Tr_type_R | Traps at the right interface | 1, donor | |
| Tr_type_B | Traps at grain boundaries and in the bulk | -1, acceptor | |
| Vdistribution | Distribution of voltages that will be simulated | 1, uniform | |
| PreCond | Use of pre-conditioner | 0, no | |
| Vscan | Direction of voltage scan | -1, down | |
| Vmin | Minimum voltage that will be simulated | 0.0 | V |
| Vmax | Maximum voltage that will be simulated | 1.4 | V |
| Vstep | Voltage step | 0.01 | V |
| until_Voc | Simulation termination at Voc | 0, no | |

## 4.4. Steady-State PL Spectra and Time-Resolved PL (TRPL)

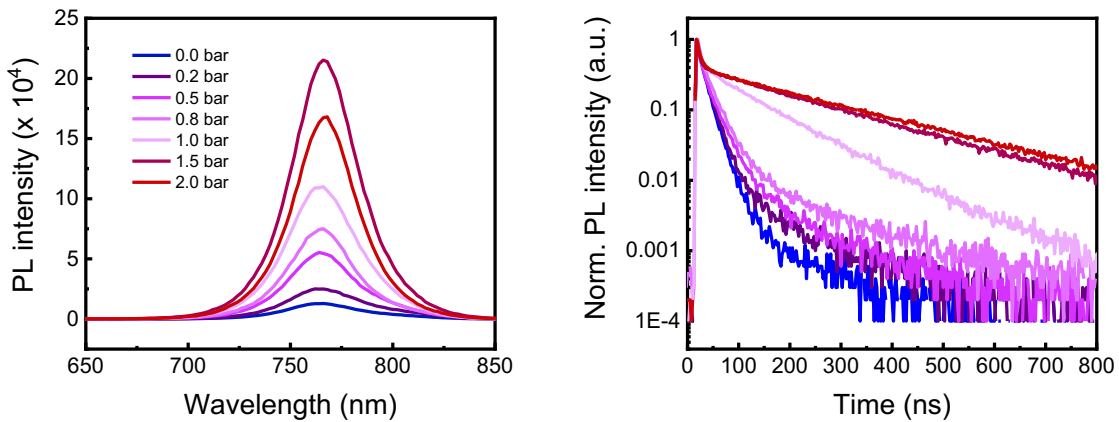

*Figure S 24: Left: Steady-state PL. Right: TRPL.*

The steady-state photoluminescence (PL) and time-resolved photoluminescence (TRPL) decay measurements were conducted to study the charge recombination in the perovskite films. As shown in Figure S 24, the enhanced PL intensity and increased average carrier lifetime are observed as increasing the evaporation rate, which suggests the reduced non-radiative recombination center in the films processed by high air flow. These results can be interpreted for the improved Voc due to low interface nonradiative recombination, which is highly consistent with the observations of the film morphology.